\documentclass[lettersize,journal]{IEEEtran}
\usepackage{aliascnt}
\usepackage{pifont}
\usepackage{mdframed}
\usepackage{fullwidth}
\usepackage[hyphens]{url}

\IEEEoverridecommandlockouts

\usepackage{amsmath,amssymb,amsfonts,amsthm}
\usepackage{algorithmic}
\usepackage[ruled,linesnumbered]{algorithm2e}
\usepackage{graphicx}
\usepackage{textcomp}
\usepackage[hidelinks]{hyperref}
\usepackage{subcaption}
\usepackage{float}
\usepackage[utf8]{inputenc}
\usepackage{tabularx}
\usepackage{booktabs}
\usepackage{float}
\usepackage{cases}
\usepackage{xcolor}
\usepackage{authblk}
\usepackage{epstopdf}
\usepackage{etoolbox}
\usepackage{stfloats}
\usepackage{breakurl}
\AtBeginEnvironment{algorithm}{\SetArgSty{textrm}}

\usepackage{caption} 

\usepackage{bm}
\def\BibTeX{{\rm B\kern-.05em{\sc i\kern-.025em b}\kern-.08em
    T\kern-.1667em\lower.7ex\hbox{E}\kern-.125emX}}

\makeatletter
\newcommand{\removelatexerror}{\let\@latex@error\@gobble}
\makeatother

\makeatletter
\newcommand{\RemoveAlgoNumber}{\renewcommand{\fnum@algocf}{\AlCapSty{\AlCapFnt\algorithmcfname}}}
\newcommand{\RevertAlgoNumber}{\algocf@resetfnum}
\makeatother

\title{Transforms for Multiplicative and Fractional Programming with Broad Applications in Edge Computing and Communication Networks}

\usepackage{graphicx}
\usepackage{mathrsfs}
\usepackage{mathtools}

\usepackage{setspace} %\doublespacing
\newtheorem{assumption}{Assumption}
\newtheorem{corollary}{Corollary}
\newtheorem{definition}{Definition}
\newtheorem{theorem}{Theorem}

\newtheorem{remark}{Remark}

\newtheorem{lemma}{Lemma}

\usepackage{enumitem}
\renewenvironment{thebibliography}[1]{
  \begin{oldthebibliography}{#1}
    \setlength{\itemsep}{0em}
    \setlength{\parskip}{0em}
}
{
  \end{oldthebibliography}
}

\begin{document}

%\author{IEEE Publication Technology,~\IEEEmembership{Staff,~IEEE,}
        % <-this % stops a space
%\thanks{This paper was produced by the IEEE Publication Technology Group. They are in Piscataway, NJ.}% <-this % stops a space
%\thanks{Manuscript received April 19, 2021; revised August 16, 2021.}}

\author{Yitong Wang, Chang Liu, Jun Zhao
        % <-this % stops a space
%\thanks{A shorter version of this work containing partial results has been published in the 2022 IEEE/ACM Symposium on Edge Computing (SEC)~\cite{9996746}. Improvements over~\cite{9996746} are explicitly discussed in this paper.}
\thanks{The authors are all with the School of Computer Science and Engineering, Nanyang Technological University (NTU), Singapore. (Contact E-mail: JunZHAO@ntu.edu.sg). A 6-page short version containing partial results is under review for the 2024 IEEE International Conference on Communication (ICC) in \cite{wang2023ICC}. IEEE Communications Society (ComSoc) writes “It is possible to submit the journal and conference version at the
same time” at \url{https://www.comsoc.org/publications/journals/ieee-transactions-wireless-communications/conference-vs-journal}}
}

%\author{\IEEEauthorblockN{Yitong Wang, Chang Liu, Jun Zhao}\\
%\IEEEauthorblockA{School of Computer Science and Engineering \\
%Nanyang Technological University, Singapore\\
%\{yitong002, liuc0063\}@e.ntu.edu.sg, junzhao@ntu.edu.sg}
%}



\maketitle

\begin{abstract}
Multiplicative Programming (MP) pertains to a spectrum of optimization problems that involve product term(s). As computational paradigms of communication systems continue to evolve, particularly concerning the offloading strategies of computationally intensive tasks simultaneously to centralized or decentralized servers, designing or optimizing effective communication systems with MP techniques becomes increasingly indispensable. Similarly, Fractional Programming (FP) is another significant branch in the optimization domain, addressing various essential scenarios in communication. For instance, in minimization optimization problems, transmission power and processing delay of communication systems are considered critical metrics. In a very recent JSAC paper by Zhao~\textit{et~al.}~\cite{zhao2023human}, an innovative transform (Zhao's Optimization Transform) was proposed for solving the minimization of MP and FP problems. Nevertheless, the resolution of optimization problems in communication systems encounters several limitations when adopting Zhao's optimization transform, especially in MP problems. Primarily, objective functions proposed in these optimization problems typically involve sum-of-products terms and the optimization variables are always discrete leading to NP-hard problems. Furthermore, multiple functions mapping to the \mbox{non-negative} domain in these scenarios can result in auxiliary variables being zero values, while the same situation is avoidable in FP problems due to the presence of these functions in the denominator. In this paper, we introduce an updated transform, building on the foundations of Zhao's original method, designed to effectively overcome these challenges by reformulating the original problem into a series of convex or concave problems. This introduced problem reformulation provides a superior iteration algorithm with demonstrable convergence to a stationary point. Additionally, we conduct an in-depth analysis of application scenarios for the proposed updated transform and provide comparative evaluations with traditional MP methodologies.
\end{abstract}
%Secondly, objective functions derived from the product of convex or concave functions mostly do not maintain convexity or concavity with respect to the optimization variables. 


\section{Overview}
%整篇的基调是为增加transform的潜力和一般性，在把proposed trasform带入到MP中时，随之而来的出现了一系列新的情况
\IEEEPARstart{F}{ractional} programming (FP), which solves the optimization problem consisting of ratio term(s), is a critical tool in the development of current communication systems and the enhancement of communication application performance. In a very recent paper accepted to JSAC, Zhao~\textit{et~al.}~\cite{zhao2023human} introduced an innovative transformation technique, referred to as \textit{Zhao's Transform}. This method is tailored to tackle the minimization of FP problems, characterized by sum-of-ratios that are commonly encountered in the design of resource allocation strategies within wireless communication networks. The previously proposed Zhao's optimization transform distinguishes itself by demonstrating superior performance when compared to previous optimization techniques, such as Dinkelbach's Method \cite{jong2012efficient}. This enhancement in performance underscores its potential effectiveness and applicability in optimizing complex communication network scenarios.

%P1: introduce the advantage of the proposed transform which can be adapted to the optimization, especially in the FP problem...

%Fractional programming 

\textit{Zhao's Transform} has great potential not only in FP problems but is equally effective in multiplicative programming (MP) problems \cite{shen2022global}. MP in wireless networks is a complex optimization technique used to enhance various aspects of system performance. This approach primarily involves solving nonlinear problems where the objective function is a product of several variables or functions, such as signal strength, bandwidth, or power levels. These optimization problems are crucial in areas like resource allocation \cite{feng2022heterogeneous}, where it is formulated to optimize the distribution of network resources for maximum utility \cite{pervezJointUserAssociation2021}, and in physical layer security, for relaying selection strategy to improve system robustness \cite{moon2019proactive}. The \mbox{non-linear} and \mbox{non-convex} properties of these problems make them challenging, necessitating advanced and heuristics algorithms, including gradient-based methods \cite{10.1007/s11276-020-02426-9}, alternative optimization (AO) approach \cite{amiriaraIRSUserAssociationIRSAided2022}, and genetic algorithms \cite{lu2022genetic}, to find optimal solutions. However, these methods still have limitations in terms of the computation complexity, convergence rate, and sensitivity of parameter modification. 

%P2: introduce the proposed transform that can be adapted to many application scenarios of MP for example...
In this paper, we extend the potential of \textit{Zhao's Transform} in the MP problems with multiple products to prove the effectiveness of this proposed transform. Unique obstacles, which would not be encountered in FP problems, are met in applying the transform to MP problems. First, there are some common scenarios related to MP problems having the special case that $B(\boldsymbol{x})=0$, where this special case can be unconsciously avoided in formulated FP problems as $B(\boldsymbol{x})$ participating in the algorithm as denominators. However, this case cannot be disregarded in the multiplicative case. Therefore, the first obstacle is introduced that auxiliary variables derived from $B(\boldsymbol{x})$ will equal zero, which conflicts with the iteration algorithms where auxiliary variables will be used as denominators. Furthermore, there are some other cases relating to discrete or mixed discrete-continuous optimization variables that result in the optimization problems as NP-hard problems, while the proposed \textit{Zhao's Transform} focuses on the continuous variables. Therefore, resolving the NP-hard problem due to discrete variables becomes another challenge.

%P3: however there are some common scenarios related to MP that have special cases like $A(\boldsymbol{x})=0$,..., so the transform needs to be updated to become more general...

%P4: also there are some cases including discrete variables that can not be taken into consideration because the transform in Part I does not include these cases (NP-hard Problems)...

In an effort to broaden the applicability of \textit{Zhao's Transform} across a wider range of wireless communication networks, particularly for the purposes of design enhancement and optimization, we introduce an \textit{Updated Transform} in this paper. This novel transform builds upon the foundational principles of the original transform in \cite{zhao2023human}. Its primary aim is to address and overcome the challenges previously identified in the context of wireless communication networks.

%过渡段
%P5: Therefore, to generalize the proposed transform and extend it to more wireless communication networks, we need to make some modifications when encountering extreme cases...

The goals of this paper are to show the scalability of \textit{Zhao's Transform} and the generalizability of \textit{Updated Transform}. The main contributions of this paper are as follows:
\begin{itemize}
    \item \textit{Scalability of Zhao's Transform:} In this paper, we review the proposed transform by Zhao \textit{et al.} in \cite{zhao2023human} and make a comprehensive conclusion of this transform in FP problems and MP problems under the assumptions of functions $A(\boldsymbol{x})$ and $B(\boldsymbol{x})$. Additionally, we describe various corollaries that emerge in different scenarios, with a particular emphasis on optimization challenges associated with sum-of-ratios and sum-of-products formulations. This comprehensive conclusion aims to illustrate the broader applicability of the former Zhao's transform proposed in diverse computational contexts.
    \item \textit{Updated Transform for the Intractable Obstacles:} We investigate the proposed updated transform in this paper and mainly focus on resolving the challenges mentioned above including $B(\boldsymbol{x})=0$ and optimization variables which are discrete or mixed discrete-continuous. After illustrating each of these challenges, we analyze the essential properties of the updated transform and further present the corollaries derived.
    \item \textit{Joint Partial Offloading Ratio and Computation Resource:} To validate the effectiveness, the proposed updated transform is first applied to the minimization of the overall computation latency of a communication network by jointly optimizing the partial offloading rate and computation frequencies of the mobile edge server and local devices (i.e., the product of the offloading ratio of computational task and the cost function of the communication system). Instead of adopting existing toolboxes to solve the problem, we obtain the optimum by adopting Karush-Kuhn-Tucker (KKT) conditions analysis and verify the generalizability of the proposed updated transform at $B(\boldsymbol{x})=0$. 
    \item \textit{Joint User Association and Resource Allocation:} We also apply the proposed updated transform into the 5G heterogeneous mobile edge computing system to make an in-depth analysis on the minimization of the energy consumption and processing delay by jointly optimizing the user association decision vector, computation offloading vector, computation resource, and transmission power. In this application scenario, we proposed a novel algorithm with the integration of the introduced updated transform and successive convex approximation (SCA) technique to tackle the NP-hard optimization problems. 
\end{itemize}

%P7: architecture of this paper 

Throughout this paper, we use $\mathbb{R}$ to denote the set of real numbers, $\mathbb{R}_+$ to denote the set of \mbox{non-negative} real numbers, and $\mathbb{R}_{++}$ to denote the set of positive real numbers. The finite-dimension vector space consisting of real $n$-dimensional vectors is denoted as $\mathbb{R}^n$. $\mathbb{N}$ is used to represent the set of natural numbers. The bold lower-case letter represents a vector. The script capital denotes a set. The bold script capital denotes a vector space in the proposed algorithm. In general, for a vector $\boldsymbol{v}$ in this paper, we denote its $i$-th dimension by $v_i$. 

%P8:In this paper, we use $\mathbb{R}$ to denote the set of real numbers, $\mathbb{R}_+$ to denote the set of nonnegative real numbers, and $\mathbb{R}_{++}$ to denote the set of positive real numbers. The finite-dimension vector space consisting of real $n$-vectors is denoted as $\mathbb{R}^n$. 

\section[Zhao's Transform]{Zhao's Transform\cite{zhao2023human}}
In this section, we have a brief review of the previously proposed \textit{Zhao's Transform} in terms of FP and MP. More details and proofs can be found in the prior paper \cite{zhao2023human}. 

\subsection{Convexity and Concavity}

In this subsection, we introduce basic definitions pertaining to convexity and concavity. To begin with,
%convex sets are defined as follows based on Section~2.1.4 of the book~\cite{citeulike:163662}.

\begin{definition}[Convex set]
A set $\mathcal{C}$ is convex if the line segment between any two points in $\mathcal{C}$ lies in $\mathcal{C}$, i.e.,
if for $\boldsymbol{x}_1, \boldsymbol{x}_2 \in \mathcal{S}$ and $t \in [0,1]$, we have $t\boldsymbol{x}_1 + (1-t)\boldsymbol{x}_2 \in \mathcal{C}$. 
\end{definition}

Let $f(\boldsymbol{x})$ be a function defined on a convex set $\mathcal{S}$, where $\mathcal{S}$ is also a subset of a real vector space. Then we have the following from (3.1) of~\cite{citeulike:163662}.

\begin{definition}[Convexity]
$f$ is convex in $\boldsymbol{x}$ if and only if for any $\boldsymbol{x}_1, \boldsymbol{x}_2 \in \mathcal{S}$ and $t \in [0,1]$, it holds that \\$f(t\boldsymbol{x}_1 + (1-t)\boldsymbol{x}_2) \leq t f(\boldsymbol{x}_1) + (1-t) f(\boldsymbol{x}_2)$.
\end{definition}

With convexity above, Lemma~\ref{lemmaConvexityvsConcavity} below presents concavity.

\begin{lemma}[Convexity versus Concavity] \label{lemmaConvexityvsConcavity}
A function $f$ is said to be concave if and only
if $-f$ is convex. 
\end{lemma}

%this paper only discusses the objective function, but actually we can also put in the constraints

\subsection{Fractional Programming}

%?? did not discuss Slater's condition

\begin{theorem}[Rigorously and formally presenting the fractional programming transform of Section~IV in~\cite{zhao2023human}] \label{thmoneratio} %which slightly generalize the results 

Let $\boldsymbol{x}$ be an \mbox{$n$-dimensional vector}. Consider functions $A(\boldsymbol{x}):\mathbb{R}^n\rightarrow \mathbb{R}_{++}$, $B(\boldsymbol{x}):\mathbb{R}^n\rightarrow \mathbb{R}_{++}$, and $E(\boldsymbol{x}):\mathbb{R}^n\rightarrow \mathbb{R}$ which all have definitions on a set $\mathcal{X}\subseteq \mathbb{R}^n$. Functions $A(\boldsymbol{x})$ and $B(\boldsymbol{x})$ are respectively convex and concave with respect to~$\boldsymbol{x}$. We aim to solve the following minimization problem~$\mathbb{P}_1$, which belongs to fractional programming with a single ratio: 
 % Given a non-empty feasible set $\mathcal{X}\subseteq \mathbb{R}^n$, a pair of positive functions that $A(\boldsymbol{x}):\mathbb{R}^n\rightarrow \mathbb{R}_{++}$ is convex and $B(\boldsymbol{x}):\mathbb{R}^n\rightarrow \mathbb{R}_{++}$ is concave, where $n\in \mathbb{N}$, and a function $E(\boldsymbol{x}):\mathbb{R}^n\rightarrow \mathbb{R}_{+}$, a minimization FP problem with a single ratio is derived as:
\begin{subequations} \label{ProbP1}
    \begin{align}
\text{Problem~$\mathbb{P}_1$: }       &\underset{\boldsymbol{x}}{\text{minimize}}\hspace{15pt} E(\boldsymbol{x})+\frac{A(\boldsymbol{x})}{B(\boldsymbol{x})} \tag{\ref{ProbP1}} \\
        &\text{subject to} \hspace{14pt}  \boldsymbol{x} \in \mathcal{X}.
    \end{align}
\end{subequations}
To solve Problem~$\mathbb{P}_1$, we will solve Problem~$\mathbb{P}_2$ below. After defining  
\begin{align}
F(\boldsymbol{x},y)\coloneqq[A(\boldsymbol{x})]^2y\!+\!\cfrac{1}{4[B(\boldsymbol{x})]^2y}, \text{~where $y \in \mathbb{R}_{++}$,} \label{eqFxy}
\end{align}
we consider
% Then this problem can be transformed equivalently to:
\begin{subequations}\label{ProbP2}
\begin{align}
\text{Problem $\mathbb{P}_2$: }        &\underset{\boldsymbol{x},y}{\text{minimize}}\hspace{15pt} E(\boldsymbol{x})+ F(\boldsymbol{x},y) \tag{\ref{ProbP2}} \\
        &\text{subject to} \hspace{14pt}  \boldsymbol{x} \in \mathcal{X} \text{ and }y \in \mathbb{R}_{++}, 
    \end{align}
\end{subequations}
The relationship between Problems $\mathbb{P}_1$ and $\mathbb{P}_2$ is as follows: If the alternating optimization method (detailed in Algorithm~\ref{algo:solveP2}) to solve Problem $\mathbb{P}_2$ converges to $(\boldsymbol{x}^*,y^*)$, then under Assumption~\ref{assumptionOptxGiveny} (presented later), $\boldsymbol{x}^*$ is a stationary point for solving Problem~$\mathbb{P}_1$.

% and $y^{(t)}\!\coloneqq\!\cfrac{1}{2A(\boldsymbol{x}^{(t)})B(\boldsymbol{x}^{(t)})}\!\in\!\mathbb{R}_{++}$, $t$ is the iteration index.
\end{theorem}

\begin{assumption} \label{assumptionOptxGiveny}
Slater's condition holds for Problem $\mathbb{P}_2(y)$ of~(\ref{ProbP2y}), where $\mathbb{P}_2(y)$ belongs to convex optimization.
Specifically, when we write out the constraint ``$\boldsymbol{x} \in \mathcal{X}$'' of~(\ref{ProbP2ycons}) according to the standard form of (4.1) in~\cite{citeulike:163662} as follows:
\begin{subnumcases}{}
\mathcal{Q}_q(\boldsymbol{x}) \leq 0, & \text{ for } $q=1,2,\ldots,Q$, \label{Qqxleq0} \\
\mathcal{R}_r(\boldsymbol{x}) = 0, & \text{ for }$r=1,2,\ldots,R$, \label{Rrxeq0}\\
\boldsymbol{x} \in \mathbb{R}^n, 
\end{subnumcases}
Slater's condition means there exists $\widehat{\boldsymbol{x}} \in \mathbb{R}^n$ such that 
\begin{subnumcases}{}
\mathcal{Q}_q(\widehat{\boldsymbol{x}}) < 0, & \text{ for } $q=1,2,\ldots,Q$, \label{Qqxleq0} \\
\mathcal{R}_r(\widehat{\boldsymbol{x}}) = 0, & \text{ for }$r=1,2,\ldots,R$.
\end{subnumcases}
In other words, for $\widehat{\boldsymbol{x}}$, (\ref{Qqxleq0}) ``\textit{strictly}'' holds while (\ref{Rrxeq0}) holds.
% In Line~\ref{OptxGivenyz} of Algorithm~\ref{algo:solveP2}, we find an optimization method to solve $\mathbb{P}_2(\boldsymbol{y}^{(j)}, \boldsymbol{z}^{(j)})$ such that the obtained $\boldsymbol{x}^{(j+1)}$ satisfies the KKT conditions of  $\mathbb{P}_2(\boldsymbol{y}^{(j)}, \boldsymbol{z}^{(j)})$.
\end{assumption}

\begin{remark}
Theorem~\ref{thmoneratio} means that to solve Problem~$\mathbb{P}_1$ which involves the fraction $\frac{A(\boldsymbol{x})}{B(\boldsymbol{x})}$, we can introduce an additional variable $y$, replace $\frac{A(\boldsymbol{x})}{B(\boldsymbol{x})}$ with $F(\boldsymbol{x},y)$ in Eq.~(\ref{eqFxy}) to obtain Problem~$\mathbb{P}_2$, and then solve Problem~$\mathbb{P}_2$ via alternating optimization (AO).
\end{remark}

\begin{remark}
The AO process to solve Problem~$\mathbb{P}_2$ means solving Problems $\mathbb{P}_2(y)$ and $\mathbb{P}_2(\boldsymbol{x})$, where $\mathbb{P}_2(y)$ is to optimize $\boldsymbol{x}$ after $y$ in Problem~$\mathbb{P}_2$ is given, and $\mathbb{P}_2(\boldsymbol{x})$ is to optimize $y$ after $\boldsymbol{x}$ in Problem~$\mathbb{P}_2$ is given. Specifically, Problems $\mathbb{P}_2(y)$ and $\mathbb{P}_2(\boldsymbol{x})$ are as follows:
\begin{subequations}\label{ProbP2y}
\begin{align}
\text{Problem $\mathbb{P}_2(y)$: }        &\underset{\boldsymbol{x}}{\text{minimize}}\hspace{15pt} E(\boldsymbol{x})+ F(\boldsymbol{x},y) \tag{\ref{ProbP2y}} \\
        &\text{subject to} \hspace{14pt}  \boldsymbol{x} \in \mathcal{X}, \label{ProbP2ycons}
    \end{align}
\end{subequations}
and
\begin{subequations}\label{ProbP2x}
\begin{align}
\text{Problem $\mathbb{P}_2(x)$: }        &\underset{y}{\text{minimize}}\hspace{15pt} E(\boldsymbol{x})+ F(\boldsymbol{x},y) \tag{\ref{ProbP2x}} \\
        &\text{subject to} \hspace{14pt}  y \in \mathbb{R}_{++}, \label{ProbP2xcons}
    \end{align}
\end{subequations}
From the expression of $F(\boldsymbol{x},y)$ in Eq.~(\ref{eqFxy}), it is straightforward to obtain that Problem $\mathbb{P}_2(x)$ just means setting $y$ as $\frac{1}{2A(\boldsymbol{x})B(\boldsymbol{x})} $. Based on the above, Algorithm~\ref{algo:solveP2} contains the AO procedure to solve Problem~$\mathbb{P}_2$. 
In Algorithm~\ref{algo:solveP2}, Line~\ref{OptxGivenyz} optimizes~$\boldsymbol{x}$ given $y$, while 
Line~\ref{OptyzGivenx} optimizes $y$ given $\boldsymbol{x}$.
\end{remark}

\makeatletter
% Remove right hand margin in algorithm
\patchcmd{\@algocf@start}% <cmd>
  {-1.5em}% <search>
  {0pt}% <replace>
  {}{}% <success><failure>
\makeatother

%\begin{minipage}{1.05\linewidth}
% \removelatexerror
 \begin{algorithm} %[H] 
%\begin{minipage}{.47\textwidth}%[width=\linewidth+2cm,leftmargin=-1cm,rightmargin=-1cm]
\caption{Solve Problem~$\mathbb{P}_{2}$ of~(\ref{ProbP2}) using alternating optimization.  Based on Theorem~\ref{thmoneratio}, under Assumption~\ref{assumptionOptxGiveny} which holds in practical problems of interest, the obtained solution for $\boldsymbol{x}$ from Algorithm~\ref{algo:solveP2} converges to a stationary point for \textbf{solving the fractional programming problem~$\mathbb{P}_1$}.}
\label{algo:solveP2}

Initialize $j \leftarrow -1$ and a feasible $\boldsymbol{x}^{(0)}$ which satisfies~Problem $\mathbb{P}_2$'s constraint(s); i.e., $\boldsymbol{x}^{(0)} \in \mathcal{X}$;

\Repeat{the relative difference between $\textrm{OBJ}_{\mathbb{P}_2}(\boldsymbol{x}^{(j)},y^{(j)})$ and $\textrm{OBJ}_{\mathbb{P}_2}(\boldsymbol{x}^{(j+1)},y^{(j+1)})$ is no greater than $\epsilon_1$ for a small positive number $\epsilon_1$ (i.e., $\frac{\textrm{OBJ}_{\mathbb{P}_2}(\boldsymbol{x}^{(j)},y^{(j)})}{\textrm{OBJ}_{\mathbb{P}_2}(\boldsymbol{x}^{(j+1)},y^{(j+1)})}- 1 \leq \epsilon_1$), where $\textrm{OBJ}_{\mathbb{P}_2}(\boldsymbol{x},y)$ denotes the objective function of~$\mathbb{P}_2$}{

Let $j \leftarrow j+1$;

With $\mathbb{P}_2(y)$ in~(\ref{ProbP2y}) denoting the problem of $\mathbb{P}_2$ given $y$, we use the Karush-Kuhn-Tucker (KKT) conditions or optimization tools like CVX~\cite{grant2009cvx} to solve $\mathbb{P}_2(y^{(j)})$, which is a convex optimization problem with $\boldsymbol{x}$ as the decision variable. Denote the obtained solution for $\boldsymbol{x}$ as $\boldsymbol{x}^{(j+1)}$; \label{OptxGivenyz} \newline \textit{//Comment: Line~\ref{OptxGivenyz} above optimizes~$\boldsymbol{x}$ given $y$. }  
% Remark~\ref{remOptxGivenyz} after Assumption~\ref{assumptionOptxGiveny} provides a discussion.

$y \leftarrow \frac{1}{2A(\boldsymbol{x}^{(j+1)})B(\boldsymbol{x}^{(j+1)})} $; \label{OptyzGivenx} \newline \textit{//Comment: Line~\ref{OptyzGivenx} above optimizes $y$ given $\boldsymbol{x}$.}

% $y_k^{(j)} \leftarrow$ for $k=1,\ldots,K$;

% $z_{\ell,q}^{(j)} \leftarrow $ for $\ell=1,2,\ldots,\Gamma$ and $q=1,2,\ldots,Q_{\ell}$; 

}
\end{algorithm} 
%\end{minipage}

\begin{remark}
Assumption~\ref{assumptionOptxGiveny} ensures that the Karush–Kuhn–Tucker (KKT) conditions are both necessary and sufficient to find the optimal solution of the convex optimization problem\footnote{Note that convex optimization may even have so solution; e.g., minimizing $x$ subject to $x \in \mathbb{R}$. However,  this paper excludes the very degenerate case where there is even no solution for the convex optimization problem $\mathbb{P}_2(y)$ since that case won't appear in practical problems that are of interest.} $\mathbb{P}_2(y)$. 
\end{remark}

This proposed transform can also be applied to the optimization problems with sum-of-ratios terms which are illustrated as follows.

\begin{corollary}[Extending Theorem~\ref{thmoneratio} to solving sum-of-ratios problems] 
Given the $K$ pairs of positive functions $A_k(\boldsymbol{x}):\mathbb{R}^n\rightarrow \mathbb{R}_{++}$ and $B_k(\boldsymbol{x}):\mathbb{R}^n\rightarrow \mathbb{R}_{++}$ where $n\in \mathbb{N}$ and $k=1,2,...,K$, and a function $E(\boldsymbol{x}):\mathbb{R}^n\rightarrow \mathbb{R}_{+}$, the minimization problem with multiple ratio terms is:
\begin{subequations} \label{ProbP3}
    \begin{align}
     \text{Problem~$\mathbb{P}_3$: }    &\underset{\boldsymbol{x}}{\text{minimize}} \hspace{15pt} E(\boldsymbol{x})+\sum_{k=1}^K \frac{A_k(\boldsymbol{x})}{B_k(\boldsymbol{x})}  \tag{\ref{ProbP3}}\\
        &\text{subject to}\hspace{14pt} \boldsymbol{x} \in \mathcal{X} .
    \end{align}
\end{subequations}
This problem can also be transformed to:
\begin{subequations}\label{ProbP4}
    \begin{align}
      \text{Problem~$\mathbb{P}_4$: }    &\underset{\boldsymbol{x},\boldsymbol{y}}{\text{minimize}} \hspace{15pt} E(\boldsymbol{x})+\sum_{k=1}^K F_k(\boldsymbol{x},y_k) \tag{\ref{ProbP4}} \\
        &\text{subject to}\hspace{14pt} \boldsymbol{x} \in \mathcal{X} \text{ and }\boldsymbol{y} \in \mathbb{R}^K,
    \end{align}
\end{subequations}
where we define  $F_k(\boldsymbol{x},y_k)$ below for each $k=1,2,\ldots,K$:
\begin{align}
    F_k(\boldsymbol{x},y_k)\coloneqq[A_k(\boldsymbol{x})]^2y_k+\frac{1}{4[B_k(\boldsymbol{x})]^2y_k},  \nonumber
\end{align}
and the collection of variables $[y_1,y_2,...,y_K]$ is denoted as $\boldsymbol{y}$ with $y_k^{(t)}\coloneqq\cfrac{1}{2A_k(\boldsymbol{x}^{(t)})B_k(\boldsymbol{x}^{(t)})}\in\mathbb{R}_{++}$, where $t$ is the iteration index.
\end{corollary}

By transforming the single-ratio (resp., multiple-ratio) problem to a series of sub-problems, the global optimum (resp., stationary point) is obtained when the function $E(\boldsymbol{x})$ is convex of the optimization variables  \cite{schaible1995fractional}. 

\subsection{Multiplicative Programming}
Based on the prior paper \cite{zhao2023human}, Zhao's optimization transform not only can be adapted to the FP but also to the MP. To further reveal the generalization of the proposed transform, we next delve into the potential of this transform to MP techniques.

\begin{theorem}[Single-Product Problem] \label{SingleMP}
    Given a \mbox{non-empty} feasible set $\mathcal{X}$, a pair of positive convex functions $A(\boldsymbol{x}):\mathbb{R}^n\rightarrow \mathbb{R}_{++}$ and $B(\boldsymbol{x}):\mathbb{R}^n\rightarrow \mathbb{R}_{++}$, where $n\in \mathbb{N}$, and a function $E(\boldsymbol{x}):\mathbb{R}^n\rightarrow \mathbb{R}_{+}$, a minimization MP with single product is:
\begin{subequations} \label{TP2}
    \begin{align} 
        &\underset{\boldsymbol{x}}{\text{minimize}} \hspace{15pt} E(\boldsymbol{x})+A(\boldsymbol{x})B(\boldsymbol{x}) \tag{\ref{TP2}} \\
        &\text{subject to} \hspace{14pt} \boldsymbol{x} \in \mathcal{X}.
    \end{align}
\end{subequations}
\end{theorem}

%\textit{Theory 2 (Single-Product Problem):} Given a non-empty feasible set $\mathcal{X}$, a pair of positive convex functions $A(\boldsymbol{x}):\mathbb{R}^n\rightarrow \mathbb{R}_{++}$ and $B(\boldsymbol{x}):\mathbb{R}^n\rightarrow \mathbb{R}_{++}$, where $n\in \mathbb{N}$, and a function $E(\boldsymbol{x}):\mathbb{R}^n\rightarrow \mathbb{R}_{+}$, a minimization MP with single product is:
%\begin{subequations}
%    \begin{align}
%        &\underset{\boldsymbol{x}}{\text{minimize}} \hspace{15pt} E(\boldsymbol{x})+A(\boldsymbol{x})B(\boldsymbol{x}) \\
%        &\text{subject to} \hspace{14pt} \boldsymbol{x} \in \mathcal{X}.
%    \end{align}
%\end{subequations}

Based on the characteristics as well as the generality of the functions (e.g. Shannon's formula and edge-device association formula) in the wireless communication system, the objective function (i.e. the product of two convex functions) does not always maintain the convexity of the optimization variables (e.g. offloading ratios and power allocations). The extension of the proposed transform is further expanded and this MP problem is reformulated as:
\begin{subequations} \label{TP2_1}
\begin{align}
    &\underset{\boldsymbol{x},y}{\text{minimize}} \hspace{15pt} E(\boldsymbol{x})+F(\boldsymbol{x},y) \tag{\ref{TP2_1}} \\
    &\text{subject to} \hspace{14pt} \boldsymbol{x} \in \mathcal{X},
\end{align}
\end{subequations}
where 
\begin{align}
    F(\boldsymbol{x},y)\coloneqq [A(\boldsymbol{x})]^2y+\frac{[B(\boldsymbol{x})]^2}{4y},\nonumber
\end{align}
and the introduced auxiliary variable $y \in \mathbb{R}_{++}$ in the $t$-iteration is equal to $y^{(t)}\coloneqq\frac{B(\boldsymbol{x}^{(t)})}{2A(\boldsymbol{x}^{(t)})}$.

\begin{corollary}[Sum-of-Products Problem] \label{MultipltMP}
    Given the $K$ pairs of positive convex functions $A_k(\boldsymbol{x}):\mathbb{R}^n\rightarrow \mathbb{R}_{++}$ and $B_k(\boldsymbol{x}):\mathbb{R}^n\rightarrow \mathbb{R}_{++}$ where $n\in \mathbb{N}$ and $k=1,2,...,K$, and a function $E(\boldsymbol{x}):\mathbb{R}^n\rightarrow \mathbb{R}_{+}$, the minimization problem with multiple product terms is:
    \begin{subequations} \label{CO2}
        \begin{align}
            &\underset{\boldsymbol{x}}{\text{minimize}} \hspace{15pt} E(\boldsymbol{x})+\sum_{k=1}^K A_k(\boldsymbol{x})B_k(\boldsymbol{x}) \tag{\ref{CO2}}\\
            &\text{subject to}\hspace{14pt} \boldsymbol{x} \in \mathcal{X},
        \end{align}
    \end{subequations}
    which is reformulated to:
    \begin{subequations} \label{CO2_2}
        \begin{align}
            &\underset{\boldsymbol{x},\boldsymbol{y}}{\text{minimize}} \hspace{15pt} E(\boldsymbol{x})+\sum_{k=1}^K F_k(\boldsymbol{x},y_k) \tag{\ref{CO2_2}} \\
            &\text{subject to}\hspace{14pt} \boldsymbol{x} \in \mathcal{X},
        \end{align}
    \end{subequations}
    where 
    \begin{align}
        F(\boldsymbol{x},y_k)\coloneqq[A_k(\boldsymbol{x})]^2y_k+\frac{[B_k(\boldsymbol{x})]^2}{4y_k},~\forall k, \nonumber
    \end{align}
    $\boldsymbol{y}$ refers to $[y_1,y_2,...,y_K]$, where $y_k^{(t)}\coloneqq\frac{B_k(\boldsymbol{x}^{(t)})}{2A_k(\boldsymbol{x}^{(t)})}\in\mathbb{R}_{++}$ and $t$ is the iteration index.
\end{corollary}

%\textit{Corollary 2 (Sum-of-Products Problem):} Given the $K$ pairs of positive convex functions $A_k(\boldsymbol{x}):\mathbb{R}^n\rightarrow \mathbb{R}_{++}$ and $B_k(\boldsymbol{x}):\mathbb{R}^n\rightarrow \mathbb{R}_{++}$ where $n\in \mathbb{N}$ and $k=1,2,...,K$, and a function $E(\boldsymbol{x}):\mathbb{R}^n\rightarrow \mathbb{R}_{+}$, the minimization problem with multiple product terms is:
%\begin{subequations}
%    \begin{align}
%        &\underset{\boldsymbol{x}}{\text{minimize}} \hspace{15pt} E(\boldsymbol{x})+\sum_{k=1}^K A_k(\boldsymbol{x})B_k(\boldsymbol{x}) \\
%        &\text{subject to}\hspace{14pt} \boldsymbol{x} \in \mathcal{X},
%    \end{align}
%\end{subequations}
%which is reformulated to:
%\begin{subequations}
%    \begin{align}
%        &\underset{\boldsymbol{x},\boldsymbol{y}}{\text{minimize}} \hspace{15pt} E(\boldsymbol{x})+\sum_{k=1}^K F_k(\boldsymbol{x},y_k) \\
%        &\text{subject to}\hspace{14pt} \boldsymbol{x} \in \mathcal{X},
%    \end{align}
%\end{subequations}
%where 
%\begin{align}
%    F(\boldsymbol{x},y_k)\coloneqq[A_k(\boldsymbol{x})]^2y_k+\frac{[B_k(\boldsymbol{x})]^2}{4y_k},~\forall k, \nonumber
%\end{align}
%$\boldsymbol{y}$ refers to $[y_1,y_2,...,y_K]$ where $y_k^{(t)}\coloneqq\cfrac{B_k(\boldsymbol{x}^{(t)})}{2A_k(\boldsymbol{x}^{(t)})}\in\mathbb{R}_{++}$ and $t$ is the iteration index.

Likewise, when function $E(\boldsymbol{x})$ exhibits convexity, it becomes feasible to ascertain the globally optimal solution and stationary point for the single-product and sum-of-products optimization problems respectively. 
A stationary point in the context of a \mbox{non-convex} problem, where the objective function may not be differentiable, is defined as a point that adheres to the generalized KKT conditions \cite{boyd2003subgradient}. In scenarios where the objective function is differentiable, the generalized KKT conditions simplify to align with the traditional KKT conditions.

%For solving optimization problems that are multidimensional and complex, this transformation approach can be effectively utilized.

%\textit{Theorem 3 (Multidimensional and Complex MP):} Given function $a(\boldsymbol{x})$

%如果赵哥没写的话，就把进一步的拓展写上
%The proposed transform possesses the potential for expansion to accommodate a broader spectrum of sum-of-product computations, as well as min-max-product issues. Furthermore, its applicability can be scaled to address problems within the realm of vector optimization. 

%\textit{Corollary 3:} Given a sequence of nondecreasing functions

\section{Updated Transform}
The core methodology proposed by Zhao~\textit{et~al.}~\cite{zhao2023human} is predominantly employed to tackle continuous problem spaces wherein the functions $A(\boldsymbol{x})$ and $B(\boldsymbol{x})$ are constricted to positive domains. This approach, however, is contingent upon specific assumptions and extant empirical scenarios, which consequentially result in the persistence of two significant yet unresolved exceptional instances. In pursuit of mitigating the constraints inherent to the aforementioned transform and to promulgate its generalizability, especially within practical sum-of-products problems, two thorny dilemmas are introduced as follows:
\begin{itemize}
    \item Dilemma 1: Problems with given functions mapped to the \mbox{non-negative} domain, i.e., function $A(\boldsymbol{x}):\mathbb{R}^n\rightarrow \mathbb{R}_+$ or $B(\boldsymbol{x}):\mathbb{R}^n\rightarrow \mathbb{R}_+$;
    \item Dilemma 2: Problems with a given discrete or mixed discrete-continuous feasible set, i.e., optimization variable $\boldsymbol{x}$ can be discrete or mixed discrete-continuous. 
\end{itemize}

For \textit{Dilemma 1}, the situation where $A(\boldsymbol{x})$ equals zero for certain values of $\boldsymbol{x}$ within the domain $\mathcal{X}$ presents a significant challenge. In such instances, the auxiliary variable approach in \cite{zhao2023human} is rendered ineffective due to the zero-denominator issue inherent in the auxiliary variable expression. A similar complication arises when $B(\boldsymbol{x})=0$, leading to the auxiliary variable $y$ becoming zero. This creates a conflict as the auxiliary variable is the denominator of the reformulated objective function (i.e., $y\neq 0$), thus resulting in the ineffective of the transformed objective function $F(\boldsymbol{x},y)$. For \textit{Dilemma 2}, dealing with discrete or mixed discrete-continuous problems, which are recognized as NP-hard problems, poses another set of difficulties. Finding an optimum solution becomes particularly challenging, as the proposed Zhao's transform is not applicable. This dilemma will be more complicated in special cases where the set $\{\boldsymbol{x} | A(\boldsymbol{x})=0\lor B(\boldsymbol{x})=0\}\subseteq\mathcal{X}$, rendering the transform process further infeasible. %Especially, case 2 gets infeasible after implementing transformations when $\{\boldsymbol{x}|A(\boldsymbol{x})=0~\text{or}~ B(\boldsymbol{x})=0\}\subseteq \mathcal{X}$. 

%To with the aforementioned transform has limitations, especially in practical sum-of-product problems

%P2: Challenges in practical cases...
%\begin{itemize}
%    \item $=0$ happens to the non-negative domain function
%    \item discrete variables in MP problems, which will be analyzed in the section-.
%\end{itemize}

To solve the dilemmas introduced in the above cases, we make updates to Zhao's optimization transform while keeping the updated transform adaptable to all mentioned problems. Firstly, we formally re-delineate the properties that the updated transform is required to have when expanding to MP and reformulating the original MP objective functions:
\begin{itemize}
    \item \textit{Decoupling} (P1): Updated transform proposed should have the same decoupling property as the proposed transform in \cite{zhao2023human} with the form $F(\boldsymbol{x},y)=F_1(A(\boldsymbol{x}))f_1(y)+F_2(B(\boldsymbol{x}))f_2(y)$, where $y$ is an auxiliary variable;
    \item \textit{Equivalent Solution} (P2): Variable $\boldsymbol{x}^*$ are the optimum of $A(\boldsymbol{x})B(\boldsymbol{x})$ if and only if $\boldsymbol{x}^*$ and $y^*$ minimize  $F(\boldsymbol{x},y)$;
    \item \textit{Relaxed Objective}
    %\footnote{In the updated transform, we relax this property and obtain the relaxation property that $\lim F(\boldsymbol{x},y^*)=A(\boldsymbol{x})B(\boldsymbol{x})$.}
    (P3): For $B(\boldsymbol{x})>0$, $F(\boldsymbol{x},y^*)=A(\boldsymbol{x})B(\boldsymbol{x})$ for variable $\boldsymbol{x}$ if and only if $y^*=\arg \min_y F(\boldsymbol{x},y)$ ; While for $B(\boldsymbol{x})=0$, $F(\boldsymbol{x},y^*)=\min_y F(\boldsymbol{x},y)$ if and only if $y^*=\arg \min_y F(\boldsymbol{x},y)$. 
    \item \textit{Convexity} (P4): Function $F(\boldsymbol{x},y)$ is convex of $y$ for fixed variable $\boldsymbol{x}$, i.e., its \textit{Hessian} is positive semidefinite;
    \item \textit{Robustness} (P5): Updated transform is not only applicable for positive-value functions which proposed in prior paper \textit{(Zhao's Transform)} but also nonnegative-value functions, i.e., $\{\boldsymbol{x}|A(\boldsymbol{x})\geq 0\lor B(\boldsymbol{x})\geq 0\}\subseteq \mathcal{X}$.   
\end{itemize}

Properties incorporated by the updated transform guarantee that the resultant optimization problem retains the same optimal solution as the original optimization problem. Regarding the objective functions' values, the relaxation to this property will be introduced and analyzed in Section~\ref{propertiesanalysis}. In order to verify the properties we deﬁne by mathematical theories, we next present and prove the pursued updated transform in detail.

%Through these properties, the transform we update ensures that the new optimization problem obtained has the same optimal solution as the original problem. For the property of the value of objective functions, we relax this property. 

\subsection{Updated Transform} \label{updatedtheorem}
\begin{theorem}[Updated Transform] \label{updatedtransformtheory}
     Given a \mbox{non-empty} feasible set $\mathcal{X}$, a pair of positive convex functions\footnote{For the case $A(\boldsymbol{x}):\mathbb{R}^n\rightarrow\mathbb{R}_{+}$ and $B(\boldsymbol{x}):\mathbb{R}^n\rightarrow\mathbb{R}_{++}$, updated transform can also be adopted to find the stationary point.} $A(\boldsymbol{x}):\mathbb{R}^n\rightarrow \mathbb{R}_{++}$ and $B(\boldsymbol{x}):\mathbb{R}^n\rightarrow \mathbb{R}_{+}$, where $n\in \mathbb{N}$, the updated transform for the single-product term is:
    \begin{align}
        F_{\text{update}}(\boldsymbol{x},y)\coloneqq[A(\boldsymbol{x})]^2(y+c)+\frac{[B(\boldsymbol{x})]^2}{4(y+c)}, \label{updatedtransform}
    \end{align}
     where $c$ is an adaptive constant derived as follows which depends on the value of $B(\boldsymbol{x})$:
    \begin{align}
        c^{(t)}=\left \{
        \begin{array}{ll}
            0, & \text{if }B(\boldsymbol{x}^{(t)})>0,\\
            c_1>0, & \text{others}, 
        \end{array}
        \right. \label{updatedconstant}
    \end{align}
   and $y\in \mathbb{R}_+$ is the introduced auxiliary variable. Specifically, 
    \begin{align}
        y^{(t)}=\left\{
        \begin{array}{ll}
            \cfrac{B(\boldsymbol{x}^{(t)})}{2A(\boldsymbol{x}^{(t)})}-c^{(t)},& \text{if }\cfrac{B(\boldsymbol{x}^{(t)})}{2A(\boldsymbol{x}^{(t)})} \geq c^{(t)}, \\
            \hspace{2pt}0, & \text{others},
        \end{array}
        \right. \label{updateauxiliary}
    \end{align}
    where $t$ is the iteration index. 
\end{theorem}

\textit{Proof:} Please see Appendix \ref{proofA}. $\hfill\blacksquare$

Based on the proof, we can further derive that when $B(\boldsymbol{x}^{(t)})\neq 0$, then the auxiliary variable $y^{(t)}=\frac{B(\boldsymbol{x}^{(t)})}{2A(\boldsymbol{x}^{(t)})}$ and constant $c^{(t)}=0$. Conversely, auxiliary variable $y^{(t)}=0$ and constant $c^{(t)}=c_1$. 

To further reduce the complexity derived from the identification of the adaptive constant and auxiliary variable, we also propose the simplified version of the updated transform to obtain the optimum of the single-product problem. 

\begin{corollary}[Simplified Transform]
    Given a \mbox{non-empty} feasible set $\mathcal{X}$, a pair of positive convex functions $A(\boldsymbol{x}):\mathbb{R}^n\rightarrow \mathbb{R}_{++}$ and $B(\boldsymbol{x}):\mathbb{R}^n\rightarrow \mathbb{R}_{+}$, where $n\in \mathbb{N}$, the updated transform for the single-product term is:
    \begin{align}
        F_{\text{update}}(\boldsymbol{x},y)\coloneqq[A(\boldsymbol{x})]^2(y+c)+\frac{[B(\boldsymbol{x})]^2}{4(y+c)},
    \end{align}
    where $c$ is a given constant and $y\in \mathbb{R}_+$ is the introduced auxiliary variable. Specifically, $y^{(t)}=\frac{B(\boldsymbol{x}^{(t)})}{2A(\boldsymbol{x}^{(t)})}$ and $t$ is the iteration index.
\end{corollary}

In order to guarantee the completeness and research-oriented of the paper, we will analyze the \textit{Updated Transform} in the following parts of this paper.

\subsection{Properties Analysis} \label{propertiesanalysis}
%P1: introduce how we can use the constant to solve the problem
The proof of Theorem~\ref{updatedtransformtheory} rigorously substantiates the applicability of the introduced auxiliary variable $y$ and constant $c$, establishing their mathematical significance. We further extended the analysis and verification of this theorem with a focus on the proposed properties. Property (P1) is strictly satisfied through the introduction of auxiliary variables; i.e., $f_1(y)=y+c$ and $f_2(y)=\frac{1}{4(y+c)}$, with the values illustrated as (\ref{updateauxiliary}). Upon defining the auxiliary variable, the surrogate function of the original objective function is obtained by defining the quadratic of each function, resulting in $F_1(A(\boldsymbol{x}))=[A(\boldsymbol{x})]^2$ and $F_2(B(\boldsymbol{x}))=[B(\boldsymbol{x})]^2$. Subsequently, we derive the second-order partial derivative of the surrogate function with respect to $y$ with $\frac{\partial^2 F_{\text{update}}}{\partial y^2}=\frac{[B(\boldsymbol{x})]^2}{2(y+c)^3}\geq 0\vspace{+1pt}$.  This outcome demonstrates that the transformed function maintains the convexity of the auxiliary variable for given variable $\boldsymbol{x}$, thereby satisfying Property (P4).

%P2: Robustness analysis
In the context of Property (P5) robustness, we specifically take the case where $A(\boldsymbol{x})$ and $B(\boldsymbol{x})$: $\mathbb{R}^n\rightarrow\mathbb{R}_{++}$ into consideration, as initially introduced in \cite{zhao2023human} \textit{(Zhao's Transform)}, then the value of $c$ can always be set as zero in each iteration for convergence. Consequently, the updated transform $F_{\text{update}}$ proposed is aligned with Zhao's optimization transform. Moreover, for the generalized case $A(\boldsymbol{x})\text{ or }B(\boldsymbol{x}):\mathbb{R}^n\rightarrow\mathbb{R}_{+}$,  our revised transform is adaptable and effective in reaching the optimum, as substantiated by the theorem's proof. Thus, we conclude that Property (P5) is satisfactorily met. %in our updated transform.

%P2-1: a paragraph specifically introduces (P3) vs. equivalent objective, and provides proof of why adding $c$ does not affect finding the optimum

In light of the detailed exploration of Property (P5), we ascertain that the equivalent objective emerges conspicuously in instances where the constant $c$ is consistently assigned the value of zero. This observation holds paramount significance, aligning with the findings elucidated in \cite{zhao2023human} of our analysis. Furthermore, the fulfillment of the equivalent solution property (P2) is corroborated within this specific context. The rigorous proof, presented in \cite{zhao2023human}, delineates the conditions under which this equivalence is maintained, thereby reinforcing the updated theorem's applicability and robustness in scenarios characterized by the stipulated value of $c$. When $c$ is set to $c_1$ in Eq.~(\ref{updatedconstant}), it leads to a critical inference regarding the transformed objective functions. Specifically, the transformed function cannot attain equivalence with the original objective function. This disparity arises because the transformed function with $B(\boldsymbol{x})=0$ cannot reach its lower bound. Nonetheless, the optimization variables can achieve equivalent values owing to the \mbox{non-decreasing} nature of $F_{\text{update}}$ within the interval $[\frac{B(\boldsymbol{x})}{2A(\boldsymbol{x})}-c,+\infty)$. Consequently, this adheres to the relaxed objective property (P3), thus enhancing the theorem's applicability and robustness property (P5). %This integral aspect underscores the theorem's potential as a foundational tool for further theoretical explorations. 

%Based on the above discussion of Property (P5), the equivalent objective property can be obtained in the former case where $c$ is always endowed with zero. Also, the equivalent solution (P3) is also satisfied in this scenario which is proved in Part I. For the 

%\textit{Proof:} $\hfill\blacksquare$

%Based on the properties (P3) of the relaxation of objective functions' values in each iteration, we focus on the proprieties (P1) and (P2) to obtain the equivalent solution after the transform. Therefore, the main considerations when $B(\boldsymbol{x})=0$ are to define the optimal solution and the interference of the current iteration with the next one. 

%P3: compare with the method adding a bias to the feasible set

%\subsection{Comparative Validity}

%By analyzing the properties of the pursued updated transform, \textit{Dilemma 1} can be solved and the equivalent optimum can still obtained. To further verify the validity and efficiency of the updated transform presented in this paper, another transform with relatively less robustness based on Zhao's optimization transform is introduced as a comparison.

%\textit{Transform with Adjusted Feasible Set:} By explaining the feasibility and analyzing this introduced transform, we take the mixed integer problem into consideration. When optimization variables include discrete variables which are denoted as $\boldsymbol{x}_1\in \{0,1\}$. Modification with  

%P4: conclusion of the proposed updated transform

\subsection{Updated Transform for Multiple-Product MP}
By analyzing the properties of Theorem~\ref{updatedtransformtheory}, it can be easy to obtain the updated transform to cater to the proposed properties aforementioned. We now apply the updated transform to general multiple-product MP problems. Given $K$ pairs of functions $A_k(\boldsymbol{x}):\mathbb{R}^n\rightarrow\mathbb{R}_{++}$ and $B_k(\boldsymbol{x}):\mathbb{R}^n\rightarrow\mathbb{R}_+$ for $k=1,2,...K$, the multiple-product MP problem is:
\begin{subequations} \label{MTP}
    \begin{align}
        &\underset{\boldsymbol{x}}{\text{minimize}} \hspace{15pt} \sum_{k=1}^K A_k(\boldsymbol{x})B_k(\boldsymbol{x}) \tag{\ref{MTP}} \label{updatedp1}\\
        &\text{subject to}\hspace{14pt} \boldsymbol{x} \in \mathcal{X}.
    \end{align}
\end{subequations}

As analysis mentioned above, Property (P2) guarantees the optimal solutions of original optimization problems are equivalent to the updated problem, though the objective functions are not equivalent as a result of Property (P3). Nevertheless, the transformed problem is:
\begin{subequations} \label{MTP_1}
    \begin{align}
        &\underset{\boldsymbol{x},\boldsymbol{y}}{\text{minimize}} \hspace{15pt} \sum_{k=1}^K F_{\text{update},k}(\boldsymbol{x},y_k) \tag{\ref{MTP_1}}\\
        &\text{subject to}\hspace{14pt} \boldsymbol{x} \in \mathcal{X},
    \end{align}
\end{subequations}
with
\begin{align}
    F_{\text{update},k}(\boldsymbol{x},y_k)\!\coloneqq\! [A_k(\boldsymbol{x}^{(t)})]^2(y_k^{(t)}
    \!\!+\!c_k^{(t)})\!+\!\frac{[B_k(\boldsymbol{x}^{(t)})]^2}{4(y_k^{(t)}\!\!+\!c_k^{(t)})}, \forall k, \nonumber
\end{align}
where we define $c_k^{(t)}$ (resp., $y_k^{(t)}$) by replacing $A$ and $B$ with $A_k$ and $B_k$ in (\ref{updatedconstant}) (resp., (\ref{updateauxiliary})). Most importantly, when $B(\boldsymbol{x})>0$ in each iteration of finding the optimum, the objective value of the transformed optimization problem is equivalent to that of the original optimization problem, i..e, property (P3) reaches the equivalent objective.

Next, to generalize the updated transform, the updated transform is further extended to the more general sum-of-functions-of-product problem. 

\begin{corollary}
    Given a sequence of \mbox{non-decreasing} functions $G_k(\cdot)$ and a sequence of products $A_k(\boldsymbol{x})B_k(\boldsymbol{x})$ where $k=1,2,...,K$, the sum-of-functions-of-product problem is:
    \begin{subequations} \label{CO4}
        \begin{align}
            &\underset{\boldsymbol{x}}{\text{minimize}}\hspace{15pt} \sum_{k=1}^K G_k(A_k(\boldsymbol{x})B_k(\boldsymbol{x})) \tag{\ref{CO4}} \label{updatedp2} \\
            &\text{subject to}\hspace{14pt} \boldsymbol{x}\in \mathcal{X}.
        \end{align}
    \end{subequations}
    
    Then, the equivalent optimum is obtained from the transformed problem by adopting the updated transform as below:
    \begin{subequations} \label{CO4_1}
        \begin{align}
            &\underset{\boldsymbol{x},\boldsymbol{y}}{\text{minimize}}\hspace{15pt} \sum_{k=1}^K G_k(F_{\text{update},k}(\boldsymbol{x},y_k)) \tag{\ref{CO4_1}} \label{corollary3} \\
            &\text{subject to}\hspace{14pt} \boldsymbol{x}\in \mathcal{X},~ \boldsymbol{y}=[y_1,...,y_k] \in \mathbb{R}_+.
        \end{align}
    \end{subequations}
\end{corollary}

\textit{Proof:} Please see Appendix \ref{proofB}. $\hfill\blacksquare$

%\textit{Corollary 4 (Multidimensional complex case):}

%P3: $f_m(\cdot)$, min-max problem

\subsection{Iterative Optimization of Multiple-products MP}
Thus far, the analysis and generalization have been presented, considering the design aspects of communication systems while validating the proposed theories and corollaries. Moving forward, we shall refocus on the sum-of-products problems, which are pivotal to the functionality of wireless communication networks, and reiterate the specific assumptions underpinning the function formulations.
\begin{itemize}
    \item Functions $A_k(\boldsymbol{x})$ and  $B_k(\boldsymbol{x})$ are all convex functions;
    \item Product of each pair $A_k(\boldsymbol{x})$ and $B_k(\boldsymbol{x})$ does not maintain the convexity;
    \item Feasible set $\mathcal{X}$ is convex and nonempty which can be expressed as the standard form.
\end{itemize}

These assumptions imply that the \mbox{non-convexity} of the sum-of-products MP problems poses considerable analytical complexity. However, for the simpler case of a single-product scenario, the \mbox{non-convexity} barrier can be surmounted more readily. By leveraging decoupling techniques, such as \textit{Logarithmic Transformation} (Page 65, Section 6.2 in \cite{horst2013handbook}), \textit{Dinkelbach's Algorithm} \cite{jong2012efficient}, and \textit{Branch and Bound (B\&B) Method} \cite{boyd2007branch}, it is feasible to converge upon a global optimum. 

To further tackle the optimization problems with multiple-products by using the updated transform, we take the scalar case of the objective functions into account and consider the problems mentioned on (\ref{updatedp1}) and (\ref{updatedp2}). Additionally, we assume that function $G_k(\cdot)$ in (\ref{updatedp2}) is not only nondecreasing but also convex. Then transformed optimization problems proposed with updated transform are all becoming convex problems. 

Therefore, we present the systematic iteration approach in the Algorithm~\ref{algorithm1} for these above \mbox{non-convex} MP problems with the introduced adaptive constant $c^{(t)}$ in (\ref{updatedconstant}) and auxiliary variables $y_k^{(t)}$ in (\ref{updateauxiliary}) of $t$-th iteration. Specifically, optimization variables and auxiliary variables can be continuously optimized until convergence by adopting a block coordinate descent algorithm with fixed values for each block variable. 

\begin{algorithm}[t]
    \renewcommand{\algorithmicrequire}{\textbf{Input:}}
    \renewcommand{\algorithmicensure}{\textbf{Output:}}
    \caption{Iteration of \mbox{Non-convex} MP Problems}
    \label{algorithm1}
    Initialize: $i=0$; optimization variable $\boldsymbol{x}^{(0)}$; 
    
    Calculate and derive auxiliary variable $\boldsymbol{y}^{(0)}$ and the adaptive constant $\boldsymbol{c}^{(0)}$ based on (\ref{updateauxiliary}) and (\ref{updatedconstant}) where $\boldsymbol{y}^{(0)}=[y_1^{(0)},...,y_K^{(0)}]$ and $\boldsymbol{c}^{(0)}=[c_1^{(0)},...,c_K^{(0)}]$;
    
    Reformulate original problem with updated transform based on Theorem~\ref{updatedtransformtheory};

    \Repeat{Convergence or reach the max iteration number $I$}{
    Obtain the optimal solution $\boldsymbol{x}^{(i+1)}$ of $(i+1)$-th iteration by solving reformulated problem when given auxiliary variable $\boldsymbol{y}^{(i)}$ and adaptive constant $\boldsymbol{c}^{(i)}$;
    
    Update $\boldsymbol{y}^{(i+1)}$ and $\boldsymbol{c}^{(i+1)}$ by (\ref{updateauxiliary}) with fixed $\boldsymbol{x}^{(i+1)}$;
    
        %\vspace{-0.4cm}
    $i\leftarrow i+1$; 
    }
\end{algorithm}

\subsection{Convergence}
To verify the convergence of the Algorithm~\ref{algorithm1}, we first can obtain via the proof in Appendix~\ref{proofA}  that the values of optimization variables are converged when fixed auxiliary variable and constant are given in each iteration as illustrated in Step 5-6. When the optimization variable is fixed, the values of the auxiliary are also converged iteratively with the given adaptive constant. without loss
of optimality, we can derive that the proposed algorithm will converge to the stationary point with less computation complexity.

\section{Application Scenarios}
In this section, we delve into the practical applications of the updated transformation, especially focusing on scenarios differentiated by the nature of the optimization variables. In cases where these variables are continuous, our analysis focuses on computation resource allocation and offloading strategies within communication systems, particularly addressing the challenges in the mobile edge computing (MEC) system \cite{mao2017mobile}. This involves optimizing resource allocation strategies in a dynamic and often constrained environment.

Furthermore, for mixed-integer optimization (MIO) scenarios, our attention shifts to user association challenges in hierarchical network structures \cite{asimMultiIRSMultiUAVAssistedMEC2023,chenEnergyEfficientUserAssociation2016,liCooperativeRechargingTransmissionStrategy2023}. Here, we investigate strategies for efficient resource and service allocation across the hierarchical network, aiming to enhance the overall network performance. 

A pivotal contribution of our work in mixed discrete-continuous cases is the development of an innovative algorithm designed to tackle \textit{Dilemma 2} instead of using conventionally alternating optimization (AO) \cite{dingJointOptimizationTransmission2022} approaches with a low convergence rate. This algorithm represents a significant step forward in solving such intricate optimization challenges where complex problems are characterized by the presence of discrete variables that render them NP-hard.
\vspace{-10pt}
\subsection{Partial Offloading with Resource Allocation}
In recent years, the strategy of partial offloading has gained significant traction in the realm of mobile computing and the Internet of Things (IoT) \cite{sisinni2018industrial}.
%This surge in interest can be attributed to the exponential growth in the capabilities of mobile devices, coupled with the increasing demand for complex, compute-intensive applications. 
Despite the advancements in hardware, the computational capacities and battery life of these mobile devices remain limited. Partial offloading \cite{saleem2020latency} emerges as a strategic solution to this limitation. It involves offloading a portion of the computational tasks from the mobile device to more powerful external servers, either in the cloud or at the edge of the network. This approach not only alleviates the computational burden on mobile devices, enhancing their performance and battery efficiency but also addresses latency-sensitive applications by leveraging the proximity and processing power of edge servers. %The integration of partial offloading into mobile and IoT systems %presents a fascinating intersection of engineering and mathematics, 
%particularly focuses on optimizing the trade-offs between local processing and remote computing and reducing the complexities of data transmission and task partitioning. 

\textbf{System Model:} To formulate the optimization problem from practical application scenarios, we consider the mobile edge computing system with $N$ user equipment (UEs) and one edge server. Assume that each UE $n$ processes the compute-intensive task that can relate to the Vehicular Edge Computing (VEC) \cite{bute2021efficient}, Metaverse \cite{zhou2023resource}, or encryption computation \cite{xiao2023joint}, and the size of the task is denoted as $C_n$. In order to enhance spectral efficiency and minimize cross-channel interference, OFDMA (Orthogonal Frequency-Division Multiple Access) is adopted to reduce inter-channel interference in this proposed system. With the computational limitation of each UE, we denote the partial rate of offloaded computation tasks as $x_n$ with $x_n\in [0,1]$. $x_n=0$ indicates the UE $n$'s task will be processed locally on the mobile device, while $x_n=1$ indicates that the computation task is totally offloaded to the edge server and processed. Specifically, when $0<x_n<1$, it indicates the proportion of tasks that are offloaded to the edge server. By denoting the number of CPU cycles required to process task one bit in the local device and edge server as 
$q_n^{\text{local}}$ and $q_n^{\text{edge}}$, we can then obtain the delay for UE $n$'s computation task at different points as
\begin{align}
    T_n=T_n^{\text{local}}+T_n^{\text{edge}}, 
\end{align}
specifically, 
\begin{align}
    &T_n^{\text{local}}=\frac{C_n^{\text{local}}q_n^{\text{local}}}{f_n^l}=\frac{(1-x_n)C_nq_n^{\text{local}}}{f_n^l},  \nonumber \\
    %&T_n^{trans}=\frac{C_n^{\text{edge}}}{\alpha_n\sqrt{p_n}}=\frac{x_nC_n}{\alpha_n\sqrt{p_n}}, \nonumber \\
    &T_n^{\text{edge}}=\frac{C_n^{\text{edge}}q_n^{\text{edge}}}{f_n^e}=\frac{x_nC_nq_n^{\text{edge}}}{f_n^e},\nonumber
\end{align}
where $f_n^l$ and $f_n^e$ represent the computation frequency of UE $n$ and edge server respectively. Compared with the latency of processing, we omit transmission latency due to the proximity of the edge server to mobile devices.

Therefore, the total energy consumption is derived as:
\begin{align}
    E_n=k_nT_n^{\text{local}}{f_n^l}^3+k_eT_n^{\text{edge}}{f_n^e}^3, 
\end{align}
where $k_n$ is the coefficient reflecting the power efficiency of UE $n$ and $k_e$ represents the analogous coefficient related to the power efficiency of the edge server. 

We next infer the cost function which jointly considers the offloading ratio and power limitation of each UE as $O_n= w_1 T_n+w_2 E_n$ and after mathematical transformation, it can be expressed as:
\begin{align}
    O_n=(1-x_n)H_{n,1}(f_n^l)+x_nH_{n,2}(f_n^e),
\end{align}
where 
\begin{align}
    &H_{n,1}(f_n^l)=C_nq_n^{\text{local}}(\frac{w_1}{f_n^l}+w_2k_n{f_n^l}^2),\nonumber\\
    &H_{n,2}(f_n^e)=C_nq_n^{\text{edge}}(\frac{w_1}{f_n^e}+w_2k_e{f_n^e}^2),\nonumber
\end{align}
and $w_1$, $w_2$ serve as weight parameters specifically designated to modulate the magnitudes of the cost components. 

\textbf{Problem Formulation:} UE $n$ can choose the offloading decision $\boldsymbol{x}\coloneqq(x_1,x_2,...,x_N)$ based on its own and the edge server's computational ability. Meanwhile, the task computation frequency locally $\boldsymbol{f^l}\coloneqq (f_1^l,f_2^l,...,f_N^l)$, the edge computation frequency $\boldsymbol{f^e}\coloneqq(f_1^e,f_2^e,...,f_N^e)$ allocated to each mobile equipment can be jointly optimized to achieve optimum. Then, the joint optimization problem incorporating energy cost expenditure, offloading determinations, and computation resources is formulated as:
\begin{subequations} \label{p1}
    \begin{align}
        \underset{\boldsymbol{x},\boldsymbol{f^l},\boldsymbol{f^e}}{\text{minimize}} ~~&\sum_{n\in \mathcal{N}} (1-x_n)H_{n,1}(f_n^l)+x_nH_{n,2}(f_n^e) \tag{\ref{p1}} \\ 
        \text{subject to } ~&~\text{($\mathbf{C}_1$):}~ 0\leq x_n \leq 1, \forall n\in \mathcal{N}, \label{p1c1}\\
        &~\text{($\mathbf{C}_2$):}~\sum\nolimits_{n=1}^N f_n^e \leq F^e,\label{p1c2}\\
        &~\text{($\mathbf{C}_3$):}~ 0\leq f_n^l \leq F_n^l, \forall n\in \mathcal{N},\label{p1c3}\\
        &~\text{($\mathbf{C}_4$):}~ 0\leq f_n^e \leq F_n^e, \forall n\in \mathcal{N},\label{p1c4}
    \end{align}
\end{subequations}
where $\mathcal{N}=\{1,2,...,N\}$ is the set of UEs. $F^e$ in $(\mathbf{C}_2)$ and $F_n^l$ in $(\mathbf{C}_3)$ indicate the maximum computation frequency of the edge server and UE $n$. $F_n^e$ in $(\mathbf{C}_4)$ represented the maximum allocated computation frequency by the edge server to UE $n$. For each constraint, $(\mathbf{C}_1)$ indicates the offloading decision and the ratio of the offloaded computation task of each UE. $(\mathbf{C}_2)$ restricts the total computation frequency of the edge server. $(\mathbf{C}_3)$ represents that the local processing frequency cannot exceed the device limit. For constraint $(\mathbf{C}_4)$, it mitigates the potential for specific computation tasks to greedily consume computational resources on the edge server. 

\textbf{Solution with Updated Transform:} To solve the above model of the practical offloading strategy, the proposed transformed in this work are adopted. The original optimization is divided into the sub-problem and the coupling of the optimization variables $\boldsymbol{x}$, $\boldsymbol{f^l}$, and $\boldsymbol{f^e}$ can be resolved by introducing auxiliary variables in (\ref{updateauxiliary}). In the $(k+1)$-th iteration, the transformed objective function is expressed as:
\begin{align}
    G(\boldsymbol{x},\boldsymbol{f^l},\boldsymbol{f^e}|\boldsymbol{u}^{(k)}\!\!,\boldsymbol{v}^{(k)}\!\!,\boldsymbol{c}^{(k)})\!=\!\!\!\underset{n\in \mathcal{N}}{\sum} g_n(x_n,f_n^l,f_n^e|u_n^{(k)}\!\!,v_n^{(k)}\!\!,c_n^{(k)}), \nonumber 
\end{align}
where $u_n^{(k)}=\frac{(1-x_n^{(k)})}{2H_{n,1}({f_n^l}^{(k)})}$ and $v_n^{(k)}=\frac{x_n^{(k)}}{2H_{n,2}({f_n^e)}^{(k)}}\in \mathbb{R}^+$ are the introduced auxiliary variables related to the convergence of the proposed algorithm when $0<x_n<1$ according to (\ref{updateauxiliary}). Function $g_n(x_n,f_n^l,f_n^e|u_n^{(k)},v_n^{(k)})$ is defined as shown in (\ref{g_n}),
%\par
%\vspace{-0.2cm}
%\begin{small}
%\begin{align}
%    g_n\hspace{-2pt}=&~[H_{n,1}(f_n^l)]^2(u_n^{(k)}+c_n^{(k)})\hspace{-2pt}+\hspace{-2pt}\frac{(1-x_n)^2}{4(u_n^{(k)}+c_n^{(k)})} \nonumber\\
%    &+\hspace{-2pt}[H_{n,2}(f_n^e)]^2(v_n^{(k)}+c_n^{(k)})\hspace{-2pt}+\hspace{-2pt}\frac{x_n^2}{4(v_n^{(k)}+c_n^{(k)})},
%\end{align}
%\end{small}
where $c_n^{(k)}$ is the introduced constant according to (\ref{updatedconstant}).

\begin{figure*}[b]
\begin{align}
    \hline \nonumber \\
    g_n=[H_{n,1}(f_n^l)]^2(u_n^{(k)}+c_n^{(k)})+\frac{(1-x_n)^2}{4(u_n^{(k)}+c_n^{(k)})}+[H_{n,2}(f_n^e)]^2(v_n^{(k)}+c_n^{(k)})+\frac{x_n^2}{4(v_n^{(k)}+c_n^{(k)})} \label{g_n}
\end{align}\vspace{-0.5cm}
\end{figure*}

Then sub-problem of the original optimization problem in the $(k+1)$-th iteration is reformulated as:
\begin{align}
    \underset{\boldsymbol{x},\boldsymbol{f^l},\boldsymbol{f^e}}{\text{minimize}} ~~~&G(\boldsymbol{x},\boldsymbol{f^l},\boldsymbol{f^e}|\boldsymbol{u}^{(k)},\boldsymbol{v}^{(k)},\boldsymbol{c}^{(k)}) \\ 
    \text{subject to } ~~&\text{(\ref{p1c1}), (\ref{p1c2}), (\ref{p1c3}), (\ref{p1c4})}.\nonumber
\end{align}

Until now, the original optimization problem can be solved iteratively, and the process of solution is listed in Algorithm \ref{offloadingdecision}. Instead of using the general convex toolboxes, we adopt Karush-Kuhn-Tucker (KKT) conditions analysis to obtain the solution to the problem in a faster way. 
\begin{algorithm}[t]
    \renewcommand{\algorithmicrequire}{\textbf{Input:}}
    \renewcommand{\algorithmicensure}{\textbf{Output:}}
    \caption{Partial Offloading Programming}
    \label{offloadingdecision}
    Initialize the index of iteration: $k=0$; optimization variable $\bm{\mathcal{X}}^{(0)}\!=\![\bm{x}^{(0)}\!,{\bm{f^l}}^{(0)},{\bm{f^e}}^{(0)}]$; 
    
    Calculate and derive auxiliary variable space: $\bm{\mathcal{A}}^{(0)}=[\bm{u}^{(0)}\!\!,\bm{v}^{(0)}]$, where $u_n^{(0)}\!=\frac{(1-x_n^{(0)})}{2H_{n,1}({f_n^l}^{(0)})}$, $v_n^{(k)}\!=\frac{x_n^{(0)}}{2H_{n,2}({f_n^e}^{(0)})}$, $\forall n\in \mathcal{N}$; Derive the constant vector: $\bm{c}^{(0)}$ based on (\ref{updatedconstant}).
    
    \Repeat{Convergence or reach the max iteration number $K$}{
    Obtain the optimal variable $\bm{\mathcal{X}}^{(k+1)}$ of $((k+1))$-th iteration by adopting the Algorithm \ref{offloadingdecisionKKT} when given auxiliary variable space $\bm{\mathcal{A}}^{(k)}$;
    
    Update $\bm{\mathcal{A}}^{(k+1)}=[\bm{u}^{(k+1)}\!\!,\bm{v}^{(k+1)}]$ and $\bm{c}^{(k+1)}$ with given $\bm{\mathcal{X}}^{(k+1)}$;
    
        %\vspace{-0.4cm}
    $k\leftarrow k+1$;
    }
\end{algorithm}

\textbf{KKT Analysis:} Before proceeding with the analysis of each KKT condition, we first derive the Lagrange function of each sub-problem by introducing multipliers for constraints as shown at the (\ref{lagrangefunction}). 
%\begin{align}
    %&L=G(\boldsymbol{x},\boldsymbol{f^l},\boldsymbol{f^e}|\boldsymbol{u}^{(k)}\!\!,\boldsymbol{v}^{(k)}\!\!,\boldsymbol{c}^{(k)})\!+\!\!\underset{n\in\mathcal{N}}{\sum}[\beta_n(-x_n)+\gamma_n(x_n\!-\!1)]\nonumber \\
    %&+\delta\cdot[(\underset{n\in\mathcal{N}}{\sum}f_n^e)-F^e]+\underset{n\in\mathcal{N}}{\sum}[\epsilon_n(-f_n^l)+\zeta_n(f_n^l-F_n^l)]+\underset{n\in\mathcal{N}}{\sum} \nonumber\\
    %&~[\eta_n(-f_n^e)+\theta_n(f_n^e-F_n^e)].
%\end{align}

\begin{figure*}[b]
\begin{small}
\begin{align}
    \hline \nonumber \\ L\!=\!G(\boldsymbol{x}\!,\boldsymbol{f^l}\!\!,\boldsymbol{f^e}|\boldsymbol{u}^{(k)}\!\!,\boldsymbol{v}^{(k)}\!\!,\boldsymbol{c}^{(k)})\!+\!\!\!\!\underset{n\in\mathcal{N}}{\sum}\![\gamma_n\hspace{-2.5pt}\cdot\hspace{-2.5pt}(x_n\!\!-\!\!1)\!-\!\beta_nx_n]\!\!+\!\delta\hspace{-2.5pt}\cdot\hspace{-2.5pt}[(\!\underset{n\in\mathcal{N}}{\sum}f_n^e)\!-\!F^e]\!+\!\!\!\!\underset{n\in\mathcal{N}}{\sum}\![\zeta_n\hspace{-2.5pt}\cdot\hspace{-2.5pt}(f_n^l\!-\!F_n^l)-\epsilon_nf_n^l]\!\!+\!\!\!\!\underset{n\in\mathcal{N}}{\sum}\![\theta_n\hspace{-2.5pt}\cdot\hspace{-2.5pt}(f_n^e\!\!-\!F_n^e)-\eta_nf_n^e] \label{lagrangefunction}
\end{align}
\end{small}
\end{figure*}

Then based on the multipliers and Lagrange function, we get KKT conditions as follows:

\noindent \textbf{Stationary:} 
\begin{subequations}\label{kktS}
\begin{align}
    &\frac{\partial L}{\partial x_n}=D_n(x_n)-\beta_n+\gamma_n=0, \\
    &\frac{\partial L}{\partial f_n^l}=Q_n(f_n^l)-\epsilon_n+\zeta_n=0,\\
    &\frac{\partial L}{\partial f_n^e}=R_n(f_n^e,\delta)-\eta_n+\theta_n=0.
\end{align}
\end{subequations}

\noindent where $D_n(x_n)=\frac{x_n-1}{2(u_n^{(k)}+c_n^{(k)})}+\frac{x_n}{2(v_n^{(k)}+c_n^{(k)})}$ relating to the variable $x_n$, $Q_n(f_n^l)=2H_{n,1}(f_n^l)H_{n,1}^{'}(f_n^l)(u_n^{(k)}+c_n^{(k)})$ relating to the variable $f_n^l$, and $R_n(f_n^e,\delta)\!=\!2H_{n,2}(f_n^e)H_{n,e}^{'}(f_n^e)(v_n^{(k)}\!+\!c_n^{(k)})+\delta$ relating to the variable $f_n^e$ and the multiplier $\delta$.  

\noindent \textbf{Complementary Slackness:} 
\begin{equation}
\begin{split}\label{kktC}
    &\text{(\ref{kktC}a): }\beta_n\cdot(-x_n)=0;\hspace{10pt} \text{(\ref{kktC}b): }\gamma_n \cdot(x_n-1)=0;\\
    &\text{(\ref{kktC}c): }\delta\cdot[(\sum\nolimits_{n\in\mathcal{N}}f_n^e)-F^e]=0;\\
    &\text{(\ref{kktC}d): }\epsilon_n\cdot(-f_n^l)=0;\hspace{12pt} \text{(\ref{kktC}e): }\zeta_n\cdot(f_n^l-F_n^l)=0; \\
    &\text{(\ref{kktC}f): }\eta_n\cdot(-f_n^e)=0;\hspace{12.5pt} \text{(\ref{kktC}g): }\theta_n\cdot(f_n^e-F_n^e)=0;
\end{split}
\end{equation}

\noindent \textbf{Primal Feasibility: } (\ref{p1c1}), (\ref{p1c2}), (\ref{p1c3}), (\ref{p1c4});

\noindent \textbf{Dual Feasibility:}
\begin{equation}
\begin{split} \label{kktD}
    \text{(\ref{kktD}a)-(\ref{kktD}e)}: \beta_n, \gamma_n, \delta,\epsilon_n, \zeta_n,\eta_n,\theta_n \geq 0, \forall n \in \mathcal{N}. 
\end{split}
\end{equation}

Under the aforementioned conditions, we then proceed to seek the optimal solutions by analyzing the KKT conditions and employing the proposed algorithm which is listed in Algorithm \ref{offloadingdecisionKKT}. 

\begin{theorem}
\label{P1Theorem}
    The optimal solution of the proposed objective function can be obtained by \textbf{Algorithm} \ref{offloadingdecisionKKT} and is expressed as:
    \begin{align}
        \left\{
        \begin{array}{l}
              x_n^*= \max\{\min \{\widetilde{x}_n,1\},0\};\\
              {f_n^l}^*= \max\{\min \{\widetilde{f}_n^l,F_n^l\},0\}; \\
              {f_n^e}^*=\max\{\min \{\widehat{f}_n^e(\delta^*),F_n^e\},0\};
        \end{array}
        \right.
    \end{align}
    where $\widetilde{x}_n$ meets the condition (\ref{kktS}a) with $D_n(x_n)|_{x_n=\widetilde{x}_n}=0$, $\widetilde{f}_n^l$ and $\widehat{f}_n^e(\delta)$ meets the condition (\ref{kktS}b) and (\ref{kktS}c) with $Q_n(f_n^l)|_{f_n^l=\widetilde{f}_n^l}=0$ and $R_n(f_n^e,\delta)|_{f_n^e=\widehat{f}_n^e(\delta)}=0$, respectively.
\end{theorem}

\textit{Proof: }Please see Appendix \ref{proofC}. $\hfill\blacksquare$

\begin{algorithm}[t]
    \renewcommand{\algorithmicrequire}{\textbf{Input:}}
    \renewcommand{\algorithmicensure}{\textbf{Output:}}
    \caption{Analyze Corresponding KKT Conditions}
    \label{offloadingdecisionKKT}
    \begin{algorithmic}[1]
        \STATE Given the auxiliary variable space $\bm{\mathcal{A}}=[\bm{u},\bm{v}]$;
        \FOR{$n\leftarrow 1 \text{ to } N$}
            \STATE Obtain $\widetilde{x}_n$ by assuming function $D_n(x_n)$ on the condition (\ref{kktS}a) equal to 0;
            \STATE Obtain $\widetilde{f}_n^l$ by assuming function $Q_n(f_n^l)$ on the condition (\ref{kktS}b) equal to 0; 
            \STATE Obtain $\widehat{f}_n^e(\delta)$ when assuming $R_n(f_n^l,\delta)=0$ in the condition (\ref{kktC}c);
        \ENDFOR $\vspace{1pt}$
        \STATE $\delta^*\!\!\leftarrow\! \!\left \{ \!\! \!
         \begin{small}
         \begin{array}{l}
            0, \text{if} \sum_{n\in\mathcal{N}}\widetilde{f}_n^e(0)\leq F^e; \\
            \text{Solution to } \sum_{n\in\mathcal{N}}\widetilde{f}_n^e(0)> F^e;
         \end{array}
         \end{small}
         \right. $
        \STATE \textbf{update} $~x_n^*$, ${f_n^l}^*$, and ${f_n^e}^*$ based on Theorem~\ref{P1Theorem};
        \RETURN The optimal variable value  $\mathcal{X}^*\!\!=[\bm{x}^*\!\!,\!{\bm{f^l}}^*\!\!,\!{\bm{f^e}}^*]$.
    \end{algorithmic}
\end{algorithm}

%\subsection{Experimental Comparison} %To implement the numerical experiment, we consider the network whose topology consists of $N=20$ mobile users and each user with a piece of mobile equipment. We set the computation frequency of the local device and the MEC server as $F^e=10$GHz and $F_n=1$GHz, respectively. The maximum allocated computation frequency by MEC server $F_n^e$ equals $2$GHz. The data size $C_n$ of the computation task in each UE is set as $250$KB and the required number of CPU cycles per bit in UE and MEC server is set to $75$cycles/bit. The computation energy efficiency coefficient $k^n$ and $k^e$ are set as $1\times 10^{-25}$. 
%To implement the numerical experiment similar to \cite{wang2023ICC}, based on the aforementioned partial offloading models, 

%\begin{figure}[t]
%    \centering
%    \includegraphics[width=0.45\textwidth,height=0.25\textwidth]{result1.eps}
%    \caption{Performance of the Joint Optimization under Different Methods}
%    \label{result1}
%\end{figure}

\vspace{-10pt}
\subsection{User Association with Resource Allocation}
%P1: HFEL 
%\cite{daiJointComputationOffloading2018,dongEnergyEfficiencyOrientedJointUser2019,fengJointTaskPartitioning2021,meiPerformanceAnalysisUser2021,qiuMultipleUAVMountedBase2020,sunJointlyOptimized3D2020,sunLocationOptimizationUser2019,sunUserAssociationResource2022,tengJointOptimizationBase2021,wuJointTrajectoryCommunication2018,yangAssociationLoadOptimization2018,yangEnergyEfficientResource2019,yinResourceAllocationBasestation2020}

After solving the practical scenarios with continuous variables, we now consider the optimization problems with mixed discrete-continuous optimization variables. In this subsection, we take the 5G heterogeneous mobile edge computing system \cite{dai2018joint,liOptimizedContentCaching2022} consisting of one Macro Base Station (MBS) \cite{wang2022resource} and one Small Base Station (SBS) \cite{qureshi2021multi} as an example and propose a novel two-tier computation offloading algorithm with the convergence of the updated transform and successive convex approximation (SCA) approach \cite{sun2016majorization}. 

\textbf{System Model:} In the proposed system, we assume each base station with one MEC server to execute the offloaded computation task, and we denote the set of base stations as $\mathcal{M}=\{1,2\}$ where $m=1$ is the SBS and $m=2$ represents the MBS. SBS is connected to the MBS via wired links. There are $N$ mobile users, denoted as $\mathcal{N}=\{1,2,...,N\}$. The SBS and mobile users are randomly distributed within the convergence of MBS. Each mobile user is connected to the SBS via wireless links and has a computation-intensive and delay-sensitive task. We define the task of each user $n$ as $D_n=(d_n,c_n)$, where $d_n$ denotes the data size of the computation task, $c_n$ is the number of CPU cycles for computing one bit of task $D_n$.

For this two-tier computation offloading framework, a mobile user can partition the task into two parts. One part is offloaded to the MEC server which is associated with the base stations, and the other part is executed locally. As the computation resource of the MEC server in SBS is limited, the part offloaded will further be partitioned into two parts when the computation resource of the MEC server on SBS is exhausted. The SBS will offload the remaining part of the task to the MEC server on the MBS. Furthermore, each mobile user can also offload the task directly to the MBS. We denote the part offloaded to the base stations as $d_{n,m}(0\leq d_{n,m}\leq d_n)$ and the part processed locally as $d_n-d_{n,m}$. In the second tier, SBS further divides the fragment $d_{n,1}$ as $d_{n,1}-d_{n,1}^{'}$ and $d_{n,1}^{'} (0\leq d_{n,1}^{'} \leq d_{n,1}\leq d_n)$, and the latter part is offloaded to the MBS. For the offloading decision of each user, we denote $x_{n,m}=\{0,1\}$ as the association variable, where $x_{n,m}=0$ if user $n\in \mathcal{N}$ is associated with base station $m\in \mathcal{M}$ and $x_{n,m}=0$ otherwise.  

For the communication model, we also adopt the OFDMA technique between mobile users and base stations, and the channel power gain is denoted as $H_{n,m}$. Then when mobile user $n$ is associated with the SBS, the signal-to-interference plus noise ratio (SINR) of the uplink will be expressed as:=
\begin{align}
    \gamma_{n,1}=\frac{P_n H_{n,1}}{\sigma^2},
\end{align}
where $P_n$ is the transmission power of mobile user $n$, and $\sigma^2$ is defined as the noise power. The inter-user interference is not taken into consideration with the adoption of OFDMA. Assuming the SBS allocates bandwidth equally to its associated users, the achievable uplink data rate of mobile user $n$ can be inferred as:
\begin{align}
    R_{n,1}=\frac{B_1}{\sum_{n=1}^Nx_{n,1}}\ln(1+\gamma_{n,1}),
\end{align}
where $B_1$ represents the bandwidth of the SBS and $\sum_{n=1}^Nx_{n,1}$ denotes the amount of mobile users associates the SBS. Similarly, if a mobile user is associated with MBS, the SINR is expressed as:
\begin{align}
    \gamma_{n,2}=\frac{P_nH_{n,2}}{\sigma^2},
\end{align}
and the achievable data rate will be:
\begin{align}
    R_{n,2}=\frac{B_2}{\sum_{n=1}^Nx_{n,2}}\ln(1+\gamma_{n,2}),
\end{align}
 where $B
_2$ are the total bandwidth of the MBS and $\sum_{n=1}^Nx_{n,2}$ are the amount of users associated with MBS.

Then we further infer the cost function of the local computation in terms of the latency and energy consumption. We denote $f_n^l$ as the computation resource when processing the task locally. Then we obtain the computation time as:
\begin{align}
    T_{n,m}^{\text{local}}=\frac{(d_n-d_{n,m})c_n}{f_n^l},
\end{align}
and the corresponding energy consumption is:
\begin{align}
    E_{n,m}^{\text{local}}=k^l(d_n-d_{n,m})c_n{f_n^l}^2,
\end{align}
where $k_n$ is the effective switched capacitance depending on the chip architecture.

For the task offloaded to the SBS, we denote the allocated computation frequency by the SBS to the task as $f_n^s$ and assume that MBS allocates fixed computation frequency $f_0$ to each offloaded task. Then the processing time in SBS is:
\begin{align}
    T_{n,1}^{\text{SBS}}=\frac{d_{n,1}}{R_{n,1}}+\frac{(d_{n,1}-d_{n,1}^{'})c_n}{f_n^s}+\frac{d_{n,1}^{'}}{r_0}+\frac{d_{n,1}^{'}c_n}{f_0},
\end{align}
where $r_0$ represents the data rate of the wired link. Then the corresponding energy consumption is derived below: \par
\vspace{-0.5cm}
\begin{small}
\begin{align}
    E_{n,1}^{\text{SBS}}\!=\!\frac{P_nd_{n,1}}{R_{n,1}}\!+\!k^S\!(d_{n,1}\!\!-\!d_{n,1}^{'}\!)c_n{f_n^s}^2\!\!+\!\!\frac{\Bar{P}d_{n,1}^{'}}{r_0}\!\!+\!\!k^Md_{n,1}^{'}c_nf_0^2,
\end{align}
\end{small}

\noindent where $\Bar{P}$ is the offloading power via wired line, $k^S$ and $k^M$ are the coefficients reflecting the power efficiency of the MEC server in SBS and MBS respectively. 

When the mobile user directly offloads the task to the MEC server in MBS, the execution time is:
\begin{align}
    T_{n,2}^{\text{MBS}}=\frac{d_{n,2}}{R_{n,2}}+\frac{d_{n,2}c_n}{f_0},
\end{align}
and the energy consumption is:
\begin{align}
    E_{n,2}^{\text{MBS}}=\frac{P_nd_{n,2}}{R_{n,2}}+k^Md_{n,2}c_nf_0^2.
\end{align}

Then the total latency and energy consumption of the task for each user can be concluded as:
\begin{align}
    &T_n=(\sum_{m\in \mathcal{M}}x_{n,m}T_{n,m}^{\text{local}})+x_{n,1}T_{n,1}^{\text{SBS}}+x_{n,2}T_{n,2}^{\text{MBS}},\\
    &E_n=(\sum_{m\in \mathcal{M}}x_{n,m}E_{n,m}^{\text{local}})+x_{n,1}E_{n,1}^{\text{SBS}}+x_{n,2}E_{n,2}^{\text{MBS}}.
\end{align}

\textbf{Problem Formulation:} We define $\boldsymbol{x}\coloneqq \{x_{n,m}\}$ as the user association decision vector, $\boldsymbol{f^l}\coloneqq \{f_n^l\}$ and $\boldsymbol{f_n^s}\coloneqq \{f_n^s\}$ as the vector of local and SBS computation frequency respectively, $\boldsymbol{P}\coloneqq [P_n|_{n\in\mathcal{N}}]$ as the transmission power vector, and $\boldsymbol{d}\coloneqq\{d_{n,m},d_{n,1}^{'}\}$ as the computation offloading vector. Then the optimization problem can be formulated as:
\begin{subequations} \label{user}
\begin{align}
     \underset{\boldsymbol{x},\boldsymbol{f^l},\boldsymbol{f^s},\boldsymbol{P},\boldsymbol{d}}{\text{minimize}} ~~&\sum_{n=1}^{N}(w_1T_n+w_2E_n) \tag{\ref{user}} \\
     \text{subject to} ~~~\hspace{2pt}& (\mathbf{C}_1) \sum_{m\in\mathcal{M}}x_{n,m}=1, \forall n\in \mathcal{N},\label{userc1} \\
     &(\mathbf{C}_2) ~~0\leq f_n^l \leq F_n, \forall n\in \mathcal{N},\label{userc2}\\
     &(\mathbf{C}_3) ~\sum_{n=1}^Nx_{n,1}f_n^s\leq F^s,\label{userc3}\\
     &(\mathbf{C}_4) ~~0\leq P_n \leq P_n^{\text{max}}, \forall n \in \mathcal{N},\label{userc4}\\
     &(\mathbf{C}_5) ~~0\leq d_{n,1}^{'} \leq d_{n,1}, \forall n\in \mathcal{N},\label{userc5}\\
     &(\mathbf{C}_6) ~~ 0\leq d_{n,m} \leq d_n, \forall n\in \mathcal{N},m\in \mathcal{M},\label{userc6}\\
     &(\mathbf{C}_7) ~~ f_n^s \geq 0, \forall n\in \mathcal{N},\label{userc7}\\
     &(\mathbf{C}_8) ~~ x_{n,m} \in \{0,1\}, \forall n\in \mathcal{N}, m\in \mathcal{M}.\label{userc8} 
\end{align}
\end{subequations}
where $w_1$ and $w_2$ represent the weight parameters designed to modulate the magnitudes of the cost components. The offloading decision constraints relate to constrains $(\mathbf{C}_1)$, $(\mathbf{C}_3)$ and $(\mathbf{C}_8)$. Constraints $(\mathbf{C}_2)$, $(\mathbf{C}_3)$, and $(\mathbf{C}_7)$ limit the frequency of computation locally and in SBS respectively. Also, constraints $(\mathbf{C}_5)$ and $(\mathbf{C}_6)$ also constructed as the limitation of computation resource. $(\mathbf{C}_4)$ limits the power resource of mobile user $n$. 

\textbf{Solution with Updated Transform and SCA:} Based on the constraints of the user association vector, we first adopt SCA technology to solve the discrete optimization variables $\boldsymbol{x}$ resulting in the problem as NP-hard. Without loss of equivalence, $(\mathbf{C}_8)$ can be rewritten as:
\begin{align}
    &~~~~x_{n,m}\in [0,1], \forall n\in \mathcal{N},m \in \mathcal{M},\label{userc8_1} \\ 
    & \text{ \textbf{and} } \sum\nolimits_{n \in \mathcal{N}}\sum\nolimits_{m\in \mathcal{M}}x_{n,m}(1-x_{n,m}) \leq 0. \label{userc8_2}
\end{align}

Note that the optimization problem has been transitioned into a continuous optimization problem, resulting in a notable reduction in computational complexity when contrasted with the direct resolution of the original discrete variable $x_{n,m}$. Nonetheless, the function $\sum\nolimits_{n \in \mathcal{N}}\sum\nolimits_{m\in \mathcal{M}}x_{n,m}(1-x_{n,m})$ in constraint (\ref{userc8_2}) is a concave function. To further facilitate the solution, we adopt a method that introduces a penalty term for this concave constraint into the objective function, which is expressed as:
\begin{align}
    \sum_{n=1}^{N}(w_1T_n+w_2E_n)-\tau \cdot\sum\nolimits_{n \in \mathcal{N}}\sum\nolimits_{m\in \mathcal{M}}x_{n,m}(x_{n,m}-1),\nonumber
\end{align}
where $\tau$ is the penalty parameter with $\tau >0$. Then the objective function becomes concave due to the concavity of the second term. Simultaneously, given the second term is differentiable, we utilize the first-order Taylor series to linearize it at each iteration. Specifically, at the $(i+1)$-th iteration\footnote{To differ the iterations of intra-sub-problem and inter-sub-problems, we symbolize the iterations of intra-sub-problem and inter-sub-problems as $k$ and $i$ respectively.}, we approximate each $\sum\nolimits_{n \in \mathcal{N}}\sum\nolimits_{m\in \mathcal{M}}x_{n,m}(x_{n,m}-1)$ with $\sum\nolimits_{n \in \mathcal{N}}\sum\nolimits_{m\in \mathcal{M}}x_{n,m}^{(i)}(x_{n,m}^{(i)}-1)+(2x_{n,m}^{(i)}-1)(x_{n,m}-x_{n,m}^{(i)})$ denoted as $H(\boldsymbol{x}|\boldsymbol{x}^{(i)})$, where $\boldsymbol{x}^{(i)}$ is the optimal solution of the $i$-th sub-problem. Therefore, the objective function is reformulated as:\par
\vspace{-0.4cm}
\begin{small}
\begin{align}
    \sum_{n=1}^{N}(w_1T_n+w_2E_n)-\tau\cdot H(\boldsymbol{x}|\boldsymbol{x}^{(i)}), \label{ofSCA}
\end{align}
\end{small}

After the mathematical transformation, we can obtain the objective function as:
\begin{align}
    \sum_{n=1}^NQ_n(\boldsymbol{x},\boldsymbol{f^l},\boldsymbol{f^s},\boldsymbol{P},\boldsymbol{d}) -\tau\cdot H(\boldsymbol{x}|\boldsymbol{x}^{(i)})
\end{align}
where $Q_n(\boldsymbol{x},\boldsymbol{f^l},\boldsymbol{f^s},\boldsymbol{P},\boldsymbol{d})=O_n(\boldsymbol{x},\boldsymbol{f^l},\boldsymbol{d})+S_n(\boldsymbol{x},\boldsymbol{P},\boldsymbol{d})+U_n(\boldsymbol{x},\boldsymbol{f^s},\boldsymbol{d})+V_n(\boldsymbol{x},\boldsymbol{d})$ and specifically,
\begin{align}
    &O_n(\boldsymbol{x},\boldsymbol{f^l},\boldsymbol{d})=\underset{m\in \mathcal{M}}{\sum}x_{n,m}(d_n-d_{n,m})(\frac{w_1}{f_n^l}+w_2k^l{f_n^l}^2)c_n,\nonumber \\
    &S_n(\boldsymbol{x},\boldsymbol{P},\boldsymbol{d})=\underset{m\in \mathcal{M}}{\sum}x_{n,m}\frac{d_{n,m}(\sum_{n=1}^N x_{n,m})}{B_m\ln(1+\frac{P_nH_{n,m}}{\sigma^2})}(w_1+w_2P_n),\nonumber \\
    &U_n(\boldsymbol{x},\boldsymbol{f^s},\boldsymbol{d})= x_{n,1}(d_{n,1}-d_{n,1}^{'})(\frac{w_1}{f_n^s}+w_2k^S{f_n^s}^2)c_n,\nonumber \\
    %&V_n(\boldsymbol{x},\boldsymbol{d})= x_{n,1}\frac{d_{n,1}^{'}}{r_0}(w_1+w_2\Bar{P}),\nonumber \\
    %&Y_n(\boldsymbol{x},\boldsymbol{d})=(x_{n,1}d_{n,1}^{'}+x_{n,2}d_{n,2})(\frac{w_1}{f_0}+w_2k^Mf_0^2)c_n.\nonumber\\
    &V_n(\boldsymbol{x},\boldsymbol{d})= x_{n,1}d_{n,1}^{'}C+x_{n,2}d_{n,2}(C-\frac{w_1+w_2\Bar{P}}{r_0}), \nonumber 
\end{align}
and the constant $C=\frac{w_1+w_2\Bar{P}}{r_0}+(\frac{w_1}{f_0}+w_2k^Mf_0^2)c_n$.

\begin{theorem}
\label{theorem5}
    The proposed objective function $Q_n(\boldsymbol{x}\!\!,\boldsymbol{f^l}\!\!,\boldsymbol{f^s}\!\!,\boldsymbol{P}\!\!,\boldsymbol{d})$ with coupled optimization variables can be transformed to the convex function with updated transform, which is expressed as $\widetilde{Q}_n(\bm{\mathcal{X}}|\bm{\mathcal{A}},\bm{\mathcal{B}})$, where $\mathcal{X}\coloneqq [\boldsymbol{x},\boldsymbol{f^l},\boldsymbol{f^s},\boldsymbol{P},\boldsymbol{d}]$ is the space of optimization variables, $\bm{\mathcal{A}}\coloneqq[\boldsymbol{\alpha},\boldsymbol{\beta},\boldsymbol{\gamma},\boldsymbol{\delta},\boldsymbol{\epsilon},\boldsymbol{\zeta},\boldsymbol{\eta},\boldsymbol{\theta}]$ is the auxiliary variables space, and $\bm{\mathcal{B}}\coloneqq [\boldsymbol{b},\boldsymbol{e},\boldsymbol{g},\boldsymbol{h},\boldsymbol{j},\boldsymbol{l},\boldsymbol{q},\boldsymbol{z}]$ is the constant space. 
\end{theorem}

%\textit{Theory 5:} The proposed objective function $Q_n(\boldsymbol{x}\!\!,\boldsymbol{f^l}\!\!,\boldsymbol{f^s}\!\!,\boldsymbol{P}\!\!,\boldsymbol{d})$ with coupled optimization variables can be transformed to the convex function with updated transform, which is expressed as $\widetilde{Q}_n(\bm{\mathcal{X}}|\bm{\mathcal{A}},\bm{\mathcal{B}})$, where $\mathcal{X}\coloneqq [\boldsymbol{x},\boldsymbol{f^l},\boldsymbol{f^s},\boldsymbol{P},\boldsymbol{d}]$ is the space of optimization variables, $\bm{\mathcal{A}}\coloneqq[\boldsymbol{\alpha},\boldsymbol{\beta},\boldsymbol{\gamma},\boldsymbol{\delta},\boldsymbol{\epsilon},\boldsymbol{\zeta},\boldsymbol{\eta},\boldsymbol{\theta}]$ is the auxiliary variables space, and $\bm{\mathcal{B}}\coloneqq [\boldsymbol{b},\boldsymbol{e},\boldsymbol{g},\boldsymbol{h},\boldsymbol{j},\boldsymbol{l},\boldsymbol{q},\boldsymbol{z}]$ is the constant space. 

\textit{Proof:} Please see Appendix \ref{th5}. $\hfill\blacksquare$

Then the transformed objective function can be derived as:\par
\vspace{-0.45cm}
\begin{small}
\begin{align}
    \sum_{n=1}^N\widetilde{Q}_n(\bm{\mathcal{X}}|\bm{\mathcal{A}},\bm{\mathcal{B}})-\tau\cdot H(\boldsymbol{x}|\boldsymbol{x}^{(i)}),
\end{align}
\end{small}

\noindent where $\widetilde{Q}_n(\bm{\mathcal{X}}|\bm{\mathcal{A}},\bm{\mathcal{B}})=\widetilde{O}_n+\widetilde{S}_n+\widetilde{U}_n+\widetilde{V}_n$ is explained in Appendix \ref{th5}.  

However, this optimization problem is not convex as the optimization variables are still coupled in the constraint (\ref{userc3}). Thus, we apply the proposed updated transform to decouple the variables as follows: \par
\vspace{-0.55cm}
\begin{small}
\begin{align}
    \sum_{n=1}^N[(f_n^s)^2(\lambda_n^{(k)}+u_n^{(k)})+\frac{x_{n,1}^2}{4(\lambda_n^{(k)}+u_n^{(k)})}]-F^s\leq 0,\label{userC3_1}\\[-20pt]\nonumber
\end{align}
\end{small}

\noindent where auxiliary variable $\lambda_n^{(k)}$ and constant $u_n^{(k)}$ obey the rules in Theorem~\ref{updatedtransformtheory}. And $\lambda_n^{(k)}=\frac{x_{n,1}^{(k)}}{2f_n^{s(k)}}$ in the $(k+1)$-th iteration. 

The the sub-problem of the original problem in the $(k+1)$-th iteration of updated transform under the $(i+1)$-th iteration of SCA approach is reformulated as:
\begin{align} \label{transformedproblem}
    \underset{\boldsymbol{x},\boldsymbol{f^l},\boldsymbol{f^s},\boldsymbol{P},\boldsymbol{d}}{\text{minimize}} ~~&\sum_{n=1}^N\widetilde{Q}_n(\bm{\mathcal{X}}|\bm{\mathcal{A}}^{(k)},\bm{\mathcal{B}}^{(k)})-\tau\cdot H(\boldsymbol{x}|\boldsymbol{x}^{(i)}) \\
    \text{subject to} ~~\hspace{2pt}& (\ref{userc1}),(\ref{userc2}),(\ref{userc4}),(\ref{userc5}),(\ref{userc6}),(\ref{userc7}),(\ref{userc8_1}),(\ref{userC3_1}) \nonumber
\end{align}

Then the original optimization problem can be addressed through an iterative process. For this user association practical application, various analysis methods can be considered. To prevent unnecessary repetition in analyses, the CVX toolboxes are utilized for solving each convex sub-problem efficiently.

\textbf{Proposed Algorithm:} In the proposed method with the convergence of SCA technique, a penalty function is employed to further facilitate the transformation of various conditions, effectively segmenting the algorithm into two complementary components: \textit{Inter-Sub-Problem Programming} and \textit{Intra-Sub-Problem Programming}. For the inter-sub-problem programming as illustrated in Algorithm \ref{inter}, it focused on deriving the optimal solution $\boldsymbol{x}^{(i)}$ in the $i$-th sub-problem. This solution is subsequently utilized in the succeeding iteration of the SCA method. Then the proposed intra-sub-problem programming as illustrated in Algorithm \ref{intra} which is obtained after adopting the updated transform will be utilized to derive the optimal solution in this succeeding iteration. Therefore, the proposed algorithm will converge and get the optimal solutions. 

\begin{algorithm}[t]
    \renewcommand{\algorithmicrequire}{\textbf{Input:}}
    \renewcommand{\algorithmicensure}{\textbf{Output:}}
    \caption{Inter-Sub-Problem Programming}
    \label{inter}
    Initialization of the iteration index: $i=1$;
    
    Initialization of the optimal solution space: $\bm{\mathcal{S}}^{(0)}\!\!\!=\!\![\bm{x}^{(0)}\!\!,{\bm{f^l}}^{(0)}\!\!,{\bm{f^s}}^{(0)}\!\!,\bm{P}^{(0)}\!\!,\bm{d}^{(0)}]$; 
    
    Adopt SCA method to obtain the problem (\ref{transformedproblem});

    \Repeat{$|\bm{\mathcal{S}}^{(i)}-\bm{\mathcal{S}}^{(i-1)}|\leq \bar{\epsilon}_0$ or reach the maximum iteration number $I$}{
    Obtain the $i$-th sub-problem by using $\bm{x}^{(i-1)}$;
    
    Solve the $i$-th sub-problem by using Algorithm \ref{intra} to get the optimal optimal variable value $\mathcal{X}^*\!=\![\bm{x}^*\!\!,{\bm{f^l}}^*\!\!,{\bm{f^s}}^*\!\!,\bm{P}^*\!\!,\bm{d}^*]$;
    
    Let $\bm{\mathcal{S}}^{i}\leftarrow \mathcal{X}^*$ of $i$-th sub-problem;
    
    Set $i\leftarrow i+1$; 
    }
    
    \textbf{return} $\bm{\mathcal{S}}^{(i)}=[\bm{x}^{(i)}\!\!,{\bm{f^l}}^{(i)}\!\!,{\bm{f^s}}^{(i)}\!\!,\bm{P}^{(i)}\!\!,\bm{d}^{(i)}]$ as the optimal solution of the original optimization problem.
\end{algorithm}\vspace{-10pt}

\begin{algorithm}[t]
    \renewcommand{\algorithmicrequire}{\textbf{Input:}}
    \renewcommand{\algorithmicensure}{\textbf{Output:}}
    \caption{Intra-Sub-Problem Programming}
    \label{intra}

    Initialization of the iteration index: $k=0$; 

    Initialization of the optimization variable space: $\bm{\mathcal{X}}^{(0)}\!=\![\bm{x}^{(0)}\!\!,{\bm{f^l}}^{(0)}\!\!,{\bm{f^s}}^{(0)}\!\!,\bm{P}^{(0)}\!\!,\bm{d}^{(0)}]$; 
    
    Calculate and derive auxiliary variable space: $\bm{\mathcal{A}}^{(0)}$ and constant space: $\bm{\mathcal{B}}^{(0)}$ based on the analysis of Theorem~\ref{theorem5};

    \Repeat{Convergence or reach the max iteration number $K$}{
    Obtain the optimal variable $\bm{\mathcal{X}}^{(k+1)}$ of $(k+1)$-th iteration by adopting CVX toolboxes when given auxiliary variable space $\bm{\mathcal{A}}^{(k)}$ and constant $\bm{\mathcal{B}}^{(k)}$;
    
    Update $\bm{\mathcal{A}}^{(k+1)}$ and $\bm{\mathcal{B}}^{(k+1)}$ with given $\bm{\mathcal{X}}^{(k+1)}$;
    
    $k\leftarrow k+1$;
    }
    \textbf{return} $\bm{\mathcal{X}}^*\leftarrow \bm{\mathcal{X}}^{(k)}$ as the optimal solution of the $i$-th sub-problem. 
\end{algorithm}

%\textbf{Experimental Comparison:}

%P2: offloading decisions

%P3: physical payer security

\section{Experimental Comparioson}
% To evaluate the performance of the proposed updated transform in the aforementioned practical scenarios, we set and adjust the parameters based on the two optimization scenarios.
In this section, we conduct experiments based on the two optimization scenarios to evaluate the performance of the proposed updated transform.
\vspace{-0.5cm}

\subsection{Scenarios 1: Partial Offloading}
Based on the partial offloading scenario, in this experiment, we consider a communication network model with the integration of AIGC technology and take the maximum diffusion steps of the reverse diffusion stage into consideration. To further greatly and clearly evaluate the performance of the proposed transform, we set the total number of users $N$ is 30. To generalize the experimental result without any distortion and error, the original value of the average error rate is set as 1, which implies the original content is complete Gaussian noise. Based on the resource limitation, the fixed discretization steps are $\Delta t_{n,0}=1/500$ and $\Delta t_{n,1}=1/1000$. The computing capacity $g_n$ and frequency of mobile user $f_n$ are set as $10$GHz and $1.5$GHz by default. The computation energy efficiency coefficient $k_n$ and $k_e$ is $10^{-26}$. For the penalty parameter, we set the value as $10^5$.

\begin{figure*}[t]
    \centering
    \begin{minipage}{0.68\textwidth}
        \centering
        \subfloat{\includegraphics[width=0.3\textwidth,height=0.28\textwidth]{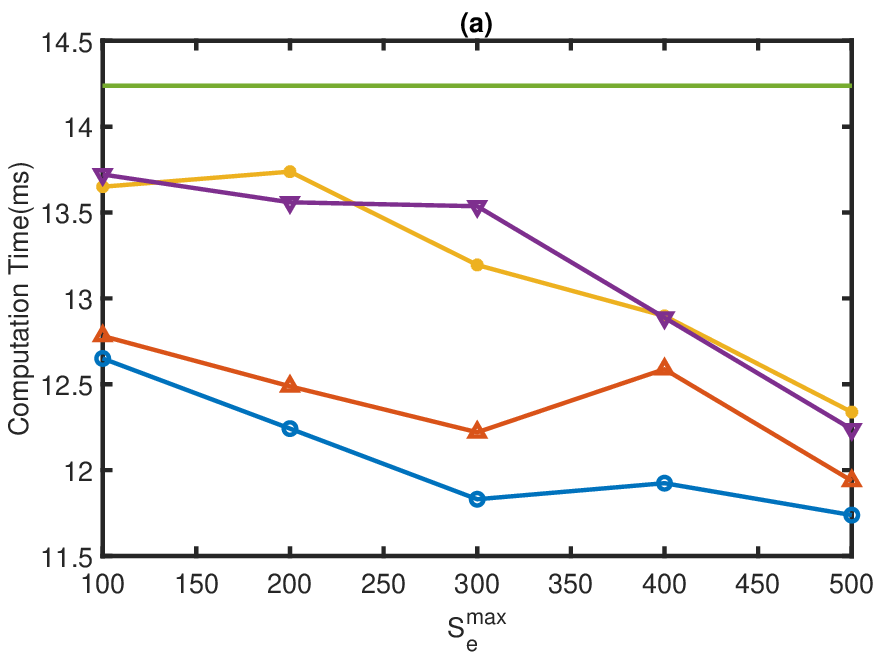}}
        \hspace{0.05in}
        \subfloat{\includegraphics[width=0.3\textwidth,height=0.28\textwidth]{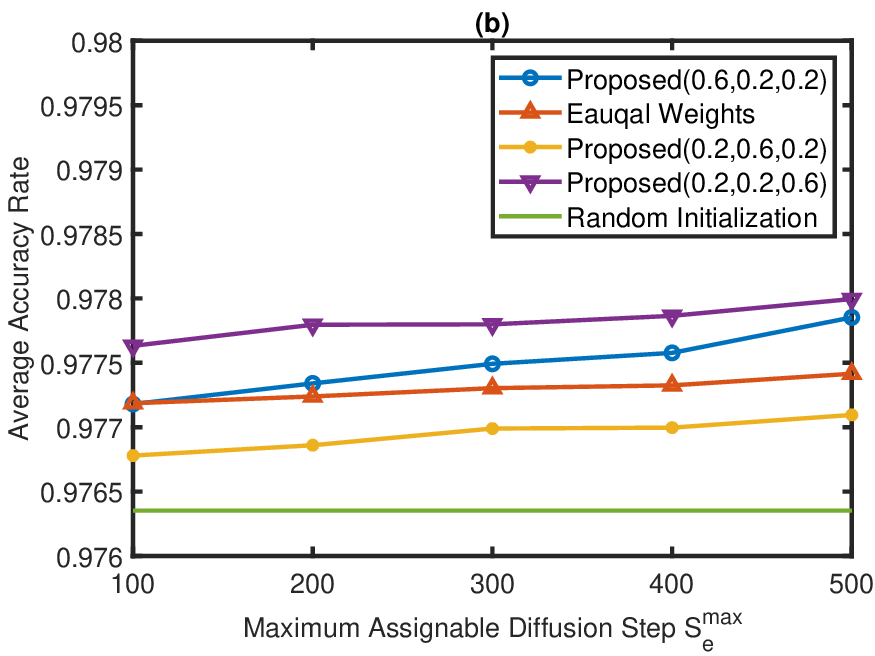}}
        \hspace{0.05in}
        \subfloat{\includegraphics[width=0.3\textwidth,height=0.28\textwidth]{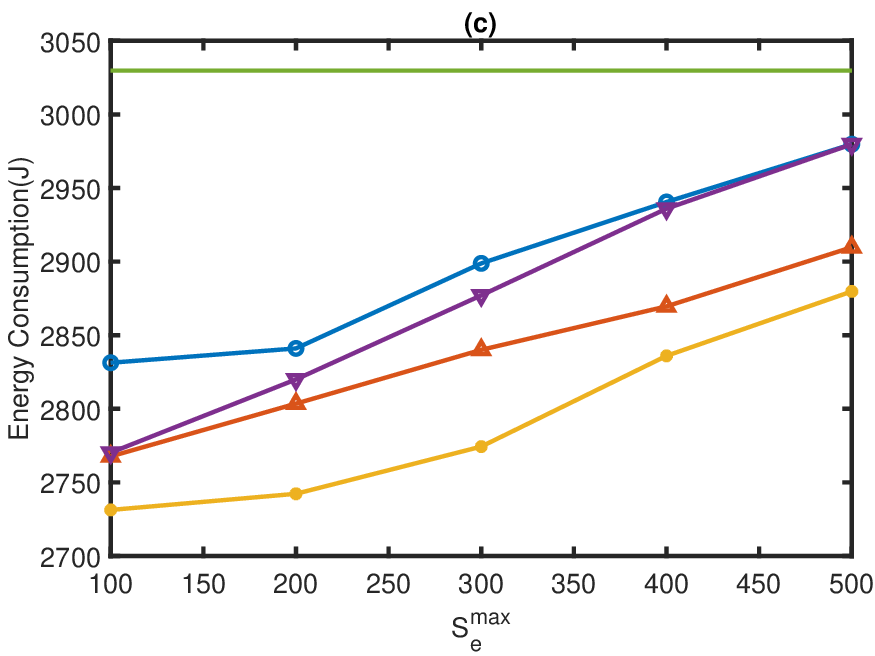}}
        %\vspace{-0.2cm}
        \caption{Consumption under Different Maximum Limitations of Resources.}\vspace{-0.9cm}
        \label{step}
    \end{minipage}
    % \hspace{0.12cm}
    \begin{minipage}{0.3\textwidth}
        \vspace{0.4cm}
        \hspace{0.21cm}
        \includegraphics[width=0.76\textwidth,height=0.63\textwidth]{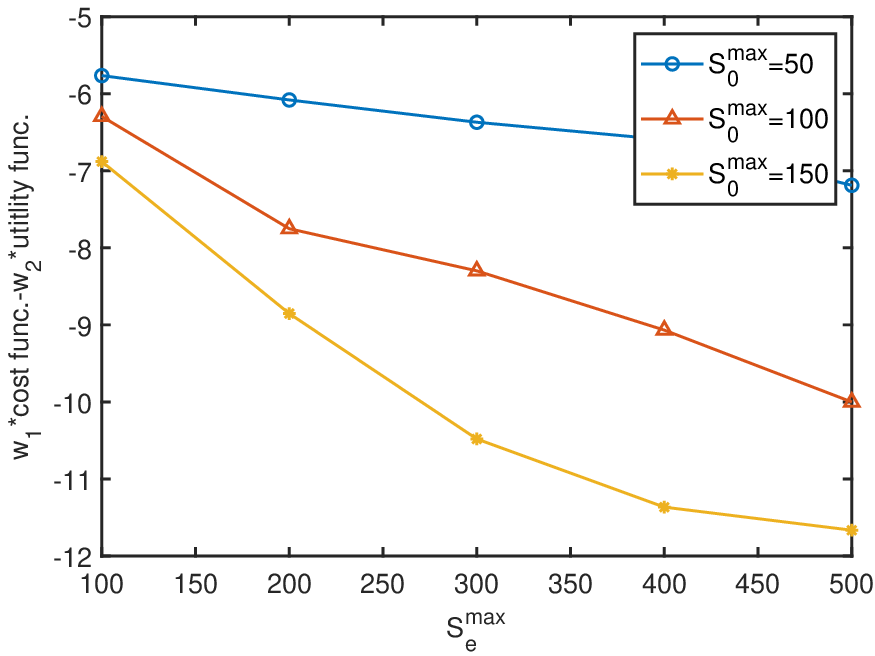}
        \vspace{-0.18cm}
        \hspace{-0.2cm}
        \caption{Joint Optimization Performance.}
        \label{performence}\vspace{-0.7cm}
    \end{minipage}
    \vspace{-3pt}
\end{figure*}

Fig.~\ref{step} and Fig.~\ref{performence} illustrate the performance of the partial offloading scenarios adopting the proposed updated transform. We can derive that the proposed updated transform method can achieve better objective function values by comparing it with diverse joint optimization scenarios. Especially, in Fig.~\ref{performence}, with the increment of the resource, all three lines exhibit an initial decline followed by a stabilization phase upon reaching an optimal solution. This observation underscores that as the system benefits from a higher resource allocation for task processing, it tends to prioritize the augmentation of the utility function to achieve superior joint optimization performance.
\vspace{-0.5cm}
\subsection{Scenarios 2: User Association}
To evaluate the performance of the proposed transform in heterogeneous communication networks, we consider a network that consists of $N=20$ users. The location of the MBS is fixed at the centre of the network. Locations of the SBS and mobile users are randomly distributed. We set the maximum transmission power of each mobile user as $100$mW. The path-loss between the MBS and the mobile user is simulated by $128.1+36.7\log_{10}(d)+\mu$ (in dB), and that between the SBS and the mobile user is set as $140.7+36.7\log_{10}(d)+\mu$ (in dB), where $d$ is the distance in km and $\mu$ satisfies the log-distribution $\mathcal{N}(0,8\text{dB})$. We set the AWGN noise power as $\sigma^2=-110$dBm. The bandwidth of MBS and SBS are set as 10MHz and 5MHz respectively. For the computation frequency of the SBS and mobile user, we set $F^s=20$GHz and $F_n=1$GHz. The data size and required number of CPU cycles per bit follow the of each computation task are set as $d_n=350$KB and $c_n=75$ cycles/bit. For the computation resource of MBS, we set the $f_0=5$GHz and $r_0=1$Gbps. The computation energy efficiency coefficient $k^l$, $k^S$, and $k^M$ are set as $1\times 10^{-25}$. For the penalty parameter $\tau$, we set the value as $10^5$. 

\begin{figure}[htbp]
    \centering
    \includegraphics[width=0.34\textwidth,height=0.22\textwidth]{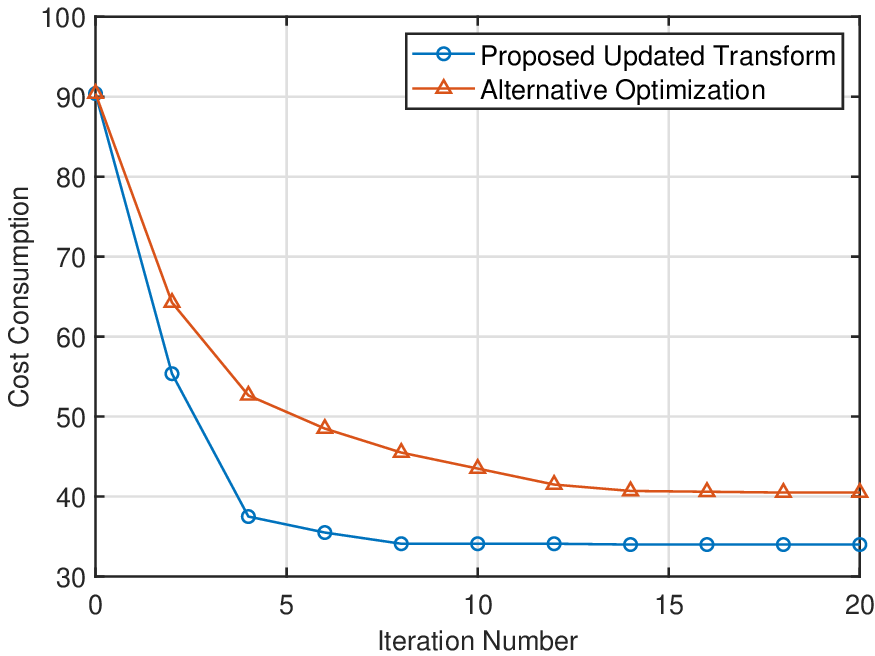}
    \vspace{-0.2cm}
    \caption{Joint Optimization Performance of Distinct Methods}
    \label{result2}
\end{figure}

To evaluate the superiority of the transform we are pursuing, we compare the optimization performance using different methods including our proposed transform and AO method. Fig.~\ref{result2} shows the result of the values of cost functions with the iterations. AO method in this proposed scenario has a larger number of iterations to converge and a worse optimization result due to the increasing computational complexity of the optimization variables and the objective function. Our proposed updated transform approach not only has a better convergence rate but also a greater optimum after iterations. Therefore, we can derive that the updated transform has a superior performance in this practical scenario based on the analysis of numerical experiments.  

\vspace{-0.5cm}

%\section{Potential Exploration}
%idea 1: modification of the constant $C$

%idea 2: find the optimum, not equivalent problem 

\vspace{-20pt}

\section{Conclusion} 
In this paper, we introduced the updated transform based on the optimization transform in prior JSAC work \cite{zhao2023human} introduced by Zhao~\textit{et~al.}, which can be adopted in the MP and FP problems to improve the communication network performance. The proposed updated transform is mainly focused on tackling the intractable dilemmas which exist in the minimization optimization problem in practical scenarios, especially mobile edge computing networks and 5G heterogeneous computing systems. By comparing with the traditional optimization techniques, e.g., alternative optimization and Newton's Method, the proposed updated transform reformulates the original optimization problem into a sequence of convex sub-problems to obtain the stationary point with less computational complexity.

\bibliographystyle{IEEEtran}
\bibliography{related.bib}

{\appendices
\section{Proof of Theorem~\ref{updatedtransformtheory}}\label{proofA}
To verify the Theorem~\ref{updatedtransformtheory}, We conduct the proof of the sufficiency and necessity with respect to the range of functions $A(\boldsymbol{x})$ and  $B(\boldsymbol{x})$. 
\subsection[]{$A(\boldsymbol{x}^{(t)})\geq 0$ and $B(\boldsymbol{x}^{(t)})>0$}
Deriving from the given adaptive constant in (\ref{updatedconstant}) and auxiliary variable in (\ref{updateauxiliary}) in the $t$-th iteration, we further obtain the exact value of $c^{(t)}$ and $y^{(t)}$ that 
\begin{align}
     y^{(t)}=\left\{
    \begin{array}{ll}
      \cfrac{B(\boldsymbol{x}^{(t)})}{2A(\boldsymbol{x}^{(t)})},& \text{if }\cfrac{B(\boldsymbol{x}^{(t)})}{2A(\boldsymbol{x}^{(t)})} \geq 0, \\
        \hspace{2pt}0, & \text{others},
    \end{array}
    \right.
\end{align}

\noindent where $c^{(t)}=0$ based on the mapped domain of variable $\boldsymbol{x}^{(t)}$ on $B(\boldsymbol{x})$. Then the update transform can be inferred upon the value of the auxiliary variable and adaptive constant:
\begin{align}
    F_{\text{update}}(\boldsymbol{x},y^{(t)})=[A(\boldsymbol{x})]^2y^{(t)}+\frac{[B(\boldsymbol{x})]^2}{4y^{(t)}},
\end{align}
which is equal to the proposed transform in \cite{zhao2023human}. Therefore, the optimum in $t$-th iteration can be obtained based on the updated transform. Thus the necessity in this case is verified. 

Next, given the updated transform in (\ref{updatedtransform}) and assuming that $y'=y+c\in \mathbb{R}_+$, the transform is derived:
\begin{align}
    F_{\text{update}}(\boldsymbol{x},y')=[A(\boldsymbol{x})]^2y'+\frac{[B(\boldsymbol{x})]^2}{4y'}.
\end{align}

Then the first-order partial derivative can be derived:
\begin{align}
    \frac{\partial F_{\text{update}}}{\partial y'^{(t)}}=[A(\boldsymbol{x})]^2-\frac{[B(\boldsymbol{x})]^2}{4y'^2}, 
\end{align}
where the stationary point can be obtained when $y'^{(t)}=\frac{B(\boldsymbol{x}^{(t)})}{2A(\boldsymbol{x}^{(t)})}$. Based on the monotonicity of the function by deriving the second-order partial derivative, we further obtain that the stationary point is the optimum in the current iteration. 

However, if the optimal solution can be always reached under the current assumptions and the robustness property of the transform is preserved by introducing the adaptive $c$ in each iteration, the constraints following need to be met:
\begin{align}
    y'^{(t)}=y^{(t)}, ~y'^{(t)}\geq 0, \text{and}~y^{(t)}\geq0,
\end{align}

Therefore, we can get that  $y^{(t)}=\frac{B(\boldsymbol{x}^{(t)})}{2A(\boldsymbol{x}^{(t)})}-c^{(t)}$ where $c^{(t)}=0$. Thus, the sufficiency and robustness (P5) of this theorem in this case are verified. 

\subsection[]{$A(\boldsymbol{x}^{(t)})\geq 0$ and $B(\boldsymbol{x}^{(t)})=0$}
In this subsection, because the property (P3) relaxed objective, we get that when $B(\boldsymbol{x})=0$, the original objective function and transformed function can not be equivalent. Therefore, what we need to verify is whether the optimal solutions between the original problem and the transformed problem are equivalent given values of $c^{(t)}$ and $y^{(t)}$.

Firstly, we focus on the monotonicity of transformed function $F_{\text{update}}(\boldsymbol{x},y)$. By taking the derivative of the transformed function and solving for stationary point based on $\frac{\partial F_{\text{update}}}{\partial y}=0$, we categorize and discuss the monotonicity of $F_{\text{update}}(\boldsymbol{x},y)$ in terms of the availability of the stationary point as follows:
\begin{itemize}
    \item \textit{Case 1:} In this case where the stationary point exists, we infer the formulation of stationary point in the $t$-th iteration as $\frac{B(\boldsymbol{x^{(t)}})}{2A(\boldsymbol{x^{(t)}})}-c^{(t)}$ with condition $\frac{B(\boldsymbol{x^{(t)}})}{2A(\boldsymbol{x^{(t)}})}\geq c^{(t)}$. Therefore, it can be derived that the transformed function is decreasing within $[0,\frac{B(\boldsymbol{x^{(t)}})}{2A(\boldsymbol{x^{(t)}})}-c^{(t)}]$ and increasing within $[\frac{B(\boldsymbol{x^{(t)}})}{2A(\boldsymbol{x^{(t)}})}-c^{(t)},+\infty)$ with fixed $x^{(t)}$. The value of $y^{(t)}$ can be further determined. 
    \item \textit{Case 2:} Instead, when the stationary point is not available, it is underlyingly implied that condition $\frac{B(\boldsymbol{x^{(t)}})}{2A(\boldsymbol{x^{(t)}})}\geq c^{(t)}$ is conflicted. Then the proposed transformed function is turned into being monotonically increasing for $y^{(t)}$ with the range $[0,+\infty)$.
\end{itemize}

The above discussion of stationary point needs to be able to both satisfy the nature of each proposed property and to bring \textit{Dilemma 1} to a resolution. Thus we further take case 2 into account and utilize the property that its feasible domain contains 0 and its \mbox{non-decreasing} nature to solve \textit{Dilemma 1}.

%这一段思路有点乱，需要作出修改
Next, we further prove the feasibility of auxiliary variable values and adaptive constant values when $B(\boldsymbol{x}^{(t)})=0$. It can be inferred that the value of $y^{(t)}$ equals zero and the value of $c^{(t)}$ is also set as $c_1>0$. The updated transform is thus able to substitute the auxiliary variables and obtain the minimum value of the transformed function based on the analysis of monotonicity in case 2. 

Until now, the theorem has been proved. We can derive that the proposed transform satisfies the properties and converges the original MP problem. $\hfill\blacksquare$

%given $c^{(t)}$ and $y^{(t)}$, verify the solution is optimum which is equal to the original solution

%given the same optimal solution, verify the value of $c^{(t)}$ and $y^{(t)}$

\section{Proof of Corollary 3} \label{proofB}
To verify this corollary, we rewrite the original problem (\ref{updatedp2}) by assuming $t_k=A_k(\boldsymbol{x})B_k(\boldsymbol{x})$ and the rewritten vision is as:
\begin{subequations}
    \begin{align}
        &\underset{\boldsymbol{x},\boldsymbol{t}}{\text{minimize}}\hspace{15pt} \sum_{k=1}^K G_k(t_k) \\
        &\text{subject to}\hspace{14pt} \boldsymbol{x}\in \mathcal{X}, \\
        &\hspace{46pt} t_k=A_k(\boldsymbol{x})B_k(\boldsymbol{x}),\forall k.
    \end{align}
\end{subequations}

Based on the property (P3), we obtain that each $t_k$ can be replaced with $\underset{y_k}{\min}[A_k(\boldsymbol{x})]^2(y_k+c_k)+\frac{[B_k(\boldsymbol{x})]^2}{4(y_k+c_k)}$ for both cases $B_k(\boldsymbol{x})=0$ and $B_k(\boldsymbol{x})\neq 0$ which has been verified in Appendix \ref{proofA}. Furthermore, as function $G_k(\cdot)$ is \mbox{non-decreasing}, the objective function can be formulated as $\underset{\boldsymbol{x}}{\min}\sum_{k=1}^K G_k(\underset{y_k}{\min}\{[A_k(\boldsymbol{x})]^2(y_k+c_k)+\frac{[B_k(\boldsymbol{x})]^2}{4(y_k+c_k)}\})$. This is equivalent to objective function in (\ref{corollary3}). Therefore, the corollary is proved and the optimum of the transformed problem is equal to the original problem. $\hfill\blacksquare$

\section{Proof of Theorem~\ref{P1Theorem}} \label{proofC}
In order to find the optimal $[\boldsymbol{x}^*,{\boldsymbol{f^l}}^*,{\boldsymbol{f^e}}^*]$ that satisfies KKT conditions, the ensuing analysis rooted in principles of optimization theory is presented. 

Deriving from the functional expression $D_n(x_n)$ as delineated in (\ref{kktS}a), two properties of this condition can be inferred contingent upon the values of auxiliary variables and Lagrange multipliers: \textit{1). $D_n(x_n)$ is \mbox{non-decreasing} for $x_n$ and 2). Specifically, we can obtain the explicit expression $\widetilde{x}_n=D_n^{-1}(0)$ by setting $\beta=0$ and $\gamma_n=0$ if $D_n(x_n)=0$ has a solution.} Then we proceed to discuss the different cases based on conditions (\ref{kktC}a) and (\ref{kktC}b) which are outlined below:
\begin{itemize}
    \item Case 1: $\widetilde{x}_n \geq 1$. In this case, we can infer that $D_n(x_n)$ is equal to or less than zero (i.e. $D_n(1)\leq 0$). Therefore, the optimal solution can be set that $\widetilde{x}_n=1$ and $\beta_n=0$ to meet the conditions, and then the value of $\gamma_n$ is equal to $-D_n(1)$ which exactly meets the condition (\ref{kktC}b).
    \item Case 2: $0< \widetilde{x}_n < 1$. We can simply set the optimal value of optimization variable $x_n$ as $\widetilde{x}_n$ with $\beta_n=0$ and $\gamma_n=0$ based on the second property.
    \item Case 3: $\widetilde{x}_n\leq 0$. Conversely with case 1, $\widetilde{x}_n\leq 0$ means when $x_n\!=\!0$, $D_n(0)$ is equal to or better than 0. Therefore, we choose the feasible solution as $\widetilde{x}_n=0$,$\gamma_n=0$, and $\beta_n=D_n(0)$. 
\end{itemize} 

Thus, we summarize all of the above cases as follows:
\begin{align}
    \left\{
    \begin{array}{l}
         x_n^*=\max\{\min \{\widetilde{x}_n,1\},0\};\\
         \beta_n^*=\max\{D_n(0),0\};\\
         \gamma_n^*=-\min\{D_n(1),0\};
    \end{array}
    \right.
\end{align}

Based on the convexity of the function $H_{n,1}(f_n^l)$, it can be inferred that the term $[H_{n,1}(f_n^l)]^2(u_n^{(k)}+c_n^{(k)})$ maintains convexity with the scalar composition theory (Page 84 in \cite{citeulike:163662}). We next obtain the \textit{\mbox{non-decreasing} nature} of the function $Q_n(f_n^l)$ as the decoupling of each optimization variable and being obtained by first-order derivation of the mentioned term. Then we can discuss the similar cases with $x_n$ by assuming $\widetilde{f}_n^l=Q_n^{-1}(0)$ and summarize the discussion as follows:
\begin{align}
    \left\{
    \begin{array}{l}
         {f_n^l}^*=\max\{\min \{\widetilde{f}_n^l,F_n^l\},0\};\\
         \epsilon_n^*=\max\{Q_n(0),0\};\\
         \zeta_n^*=-\min\{Q_n(F_n^l),0\};
    \end{array}
    \right.
\end{align}

According to the analysis of the optimization variable $f_n^l$, we obtain similar characteristics of the variable $f_n^e$ in the function $R_n(f_n^e,\delta)$. Assume that when $R_n(f_n^e,\delta)=0$, the value expression of $f_n^e$ is denoted as $\widehat{f}_n^e(\delta)$. We discuss the cases based on the (\ref{kktS}c): 
\begin{itemize}
    \item Case 1: $\widehat{f}_n^e(\delta) \geq F_n^e$. In this case, we can infer that $R_n(F_n^e,\delta)\leq 0$). Therefore, the optimal solution can be set that $\widehat{f}_n^e(\delta) = F_n^e$ and $\eta_n=0$ to meet the conditions, and then the value of $\theta_n$ is $-R_n(F_n^e,\delta)$ which exactly meets the conditions.
    \item Case 2: $0< \widehat{f}_n^e(\delta) < 1$. We can set the optimal value as $\widehat{f}_n^e(\delta)$ with $\eta_n=0$ and $\theta_n=0$ based on the second property.
    \item Case 3: $\widehat{f}_n^e(\delta)\leq 0$. Unlike case 1, $\widehat{f}_n^e(\delta)\leq 0$ means when $\widehat{f}_n^e(\delta)\!=\!0$, $R_n(0,\delta)$ is equal to or better than 0. Therefore, we choose the feasible solution as $\widehat{f}_n^e(\delta)=0$, $\theta_n=0$, and $\eta_n=R_n(0,\delta)$. 
\end{itemize} 

Therefore, for the above analysis, we summarise that:
\begin{align}
    \left\{
    \begin{array}{l}
         \widetilde{f}_n^e(\delta)=\max\{\min \{\widehat{f}_n^e(\delta),F_n^e\},0\};\\
         \widetilde{\eta}_n(\delta)=\max\{R_n(0,\delta),0\};\\
         \widetilde{\theta}_n(\delta)=-\min\{R_n(F_n^e,\delta),0\};
    \end{array}
    \right.
\end{align}

For the optimization variable $f_n^e$, we cannot directly obtain the feasible solution while getting the formulation related to the multiplier $\delta$. Drawing from the condition (\ref{kktC}c), we discuss the value of $\delta$ based on the scenarios assuming $\delta=0$:
\begin{itemize}
    \item Case 1: $\sum_{n\in\mathcal{N}}\widetilde{f}_n^e(0)\leq F^e$. In this situation, we set $\delta=0$ to meet the condition (\ref{kktD}c). 
    \item Case 2: $\sum_{n\in\mathcal{N}}\widetilde{f}_n^e(0)> F^e$. Conversely, we need to find the optimal value denoted as $\widetilde{\delta}$ of the $\delta$ based on the condition (\ref{kktC}c) by adopting the bisection method.
\end{itemize}

Consequently, the feasible solution of $\delta$ is expressed as:
\begin{align}
    \delta^*=\left\{
    \begin{array}{ll}
        0, &  \sum_{n\in\mathcal{N}}\widetilde{f}_n^e(0)\leq F^e,\\
        \widetilde{\delta}, & \text{others}.
    \end{array}
    \right.
\end{align}

Until now, the optimal solution of the optimization variables $[\boldsymbol{x}^*\!\!,\!{\boldsymbol{f^l}}^*\!\!,\!{\boldsymbol{f^e}}^*]$ and Lagrange multiplier $[\boldsymbol{\beta}^*\!,\!\boldsymbol{\gamma}^*\!,\!\delta^*\!,\!\boldsymbol{\epsilon}^*\!,\!\boldsymbol{\zeta}^*\!,\!\boldsymbol{\eta}^*\!,\!\boldsymbol{\theta}^*]$ can be obtained by above analysis. $\hfill\blacksquare$

\section{Proof of Theorem~\ref{theorem5}} \label{th5}
To decouple the optimization variables in each term of the objective function, we adopt the updated transform and introduce auxiliary variables in $(k+1)$-th iteration as follows:

For the function $O_n(\boldsymbol{x},\boldsymbol{f^l},\boldsymbol{d})$, we introduce the auxiliary variables $\boldsymbol{\alpha}$ and $\boldsymbol{\beta}$. Then the function in $(k+1)$-th iteration can be transformed into $\widetilde{O}_n(\boldsymbol{x},\boldsymbol{f^l},\boldsymbol{d}|\boldsymbol{\alpha}^{(k)},\boldsymbol{\beta}^{(k)},\boldsymbol{b}^{(k)},\boldsymbol{e}^{(k)})$ which equals to:
\begin{align}
    \sum_{m\in\mathcal{M}}[\widehat{O}_{n,m}(\boldsymbol{f^l}\!\!,\boldsymbol{d}|\boldsymbol{\beta}^{(k)}\!\!,\boldsymbol{e}^{(k)})]^2(\alpha_{n,m}^{(k)}\!+\!b_{n,m}^{(k)})\!+\!\frac{x_{n,m}^2c_n}{4(\alpha_{n,m}^{(k)}\!+\!b_{n,m}^{(k)})},\nonumber
\end{align}

\noindent where $\widehat{O}_{n,m}(\boldsymbol{f^l}\!\!,\boldsymbol{d}|\boldsymbol{\beta}^{(k)}\!\!,\boldsymbol{e}^{(k)})\!\!=\!\!(\frac{w_1}{f_n^l}\!+\!w_2k^l{f_n^l}^2)^2(\beta_{n,m}^{(k)}+e_{n,m}^{(k)})+\frac{(d_n-d_{n,m})^2}{4(\beta_{n,m}^{(k)}+e_{n,m}^{(k)})}$. When $x_{n,m}\neq 0$ and $d_n-d_{n,m} \neq 0$, values of auxiliary variables are: 
\begin{align}
    &\beta_{n,m}^{(k)}=\frac{d_n-d_{n,m}^{(k)}}{2[\frac{w_1}{{f_n^l}^{(k)}}+w_2k^l({f_n^l}^{(k)})^2]},\nonumber \\
    &\alpha_{n,m}^{(k)}=\frac{x_{n,m}^{(k)}c_n}{2\widehat{O}_{n,m}({\boldsymbol{f^l}}^{(k)},\boldsymbol{d}^{(k)}|\boldsymbol{\beta}^{(k)},\boldsymbol{e}^{(k)})}, \nonumber
\end{align}
and the selection of constant $\boldsymbol{b}^{(k)}$ and $\boldsymbol{e}^{(k)}$ follows the rule in (\ref{updatedconstant}) and the following constants obey the same rule. 

For function $S_n(\boldsymbol{x},\boldsymbol{P},\boldsymbol{d})$, it can be transformed as $\widetilde{S}_n(\boldsymbol{x},\boldsymbol{P},\boldsymbol{d}|\boldsymbol{\gamma}^{(k)},\boldsymbol{\delta}^{(k)},\boldsymbol{g}^{(k)},\boldsymbol{h}^{(k)})$ which equals to:
\begin{align}
    \sum_{m\in\mathcal{M}}[\widehat{S}_{n,m}(\boldsymbol{P}\!\!,\boldsymbol{d}|\boldsymbol{\delta}^{(k)}\!\!,\boldsymbol{h}^{(k)})]^2(\gamma_{n,m}^{(k)}\!+\!g_{n,m}^{(k)})\!+\!\frac{x_{n,m}^2}{4(\gamma_{n,m}^{(k)}\!+\!g_{n,m}^{(k)})},\nonumber
\end{align}
where 
\begin{align}
    &\widehat{S}_{n,m}(\boldsymbol{P}\!\!,\boldsymbol{d}|\boldsymbol{\delta}^{(k)}\!\!,\boldsymbol{h}^{(k)})\!\!=\!\![\overline{S}(P_n)]^2(\delta_{n,m}^{(k)}\!\!+\!\!h_{n,m}^{(k)})\!\!+\!\!\frac{d_{n,m}^2}{4(\delta_{n,m}^{(k)}+h_{n,m}^{(k)})},\nonumber \\
    &\hspace{60pt}\overline{S}(P_n)=\frac{w_1+w_2P_n}{B_m\ln(1+\frac{P_nH_{n,m}}{\sigma^2})}, \nonumber
\end{align}
$\boldsymbol{g}^{(k)},\boldsymbol{h}^{(k)}$ are the constant and auxiliary variables are expressed as:
\begin{align}
    &\delta_{n,m}^{(k)}=\frac{d_{n,m}^{(k)}}{2\overline{S}(P_n^{(k)})},\nonumber \\
    &\gamma_{n,m}^{(k)}=\frac{x_{n,m}^{(k)}}{2\widehat{S}_{n,m}(\boldsymbol{P}^{(k)},\boldsymbol{d}^{(k)}|\boldsymbol{\delta}^{(k)},\boldsymbol{h}^{(k)})},\nonumber
\end{align}
when the constants are equal to zero.

For function $U_n(\boldsymbol{x},\boldsymbol{f^s},\boldsymbol{d})$, we introduce the auxiliary $\boldsymbol{\epsilon}$ and $\boldsymbol{\zeta}$ to transform it to $\widetilde{U}_n(\boldsymbol{x},\boldsymbol{f^s},\boldsymbol{d}|\boldsymbol{\epsilon}^{(k)},\boldsymbol{\zeta}^{(k)},\boldsymbol{j}^{(k)},\boldsymbol{l}^{(k)})$ being equal to:
\begin{align}
    [\widehat{U}_n(\boldsymbol{f^s},\boldsymbol{d}|\boldsymbol{\zeta}^{(k)},\boldsymbol{l}^{(k)})]^2(\epsilon_n^{(k)}+j_n^{(k)})+\frac{x_{n,1}^2}{4(\epsilon_n^{(k)}+j_n^{(k)})},\nonumber
\end{align}
where the constants are denoted as $\boldsymbol{j}^{(k)}$ and $\boldsymbol{l}^{(k)}$. Function $\widehat{U}_n$ is expressed as:
\begin{align}
    \widehat{U}_n(\boldsymbol{f^s}\!\!,\boldsymbol{d}|\boldsymbol{\zeta}^{(k)}\!\!,\boldsymbol{l}^{(k)})\!\!=\!\!(\frac{w_1}{f_n^s}\!\!+\!\!w_2k^S{f_n^s}^2)^2(\zeta_n^{(k)}\!\!\!+\!\!l_n^{(k)})\!\!+\!\!\frac{(d_{n,1}\!\!-\!d_{n,1}^{'})^2}{4(\zeta_n^{(k)}\!\!\!+\!l_n^{(k)})}. \nonumber
\end{align}

Based on the (\ref{updateauxiliary}), if constants are equal to 0, auxiliary variables can be inferred as:
\begin{align}
    &\epsilon_n^{(k)}= \frac{x_{n,1}^{(k)}}{2\widehat{U}_n({\boldsymbol{f^s}}^{(k)},\boldsymbol{d}^{(k)}|\boldsymbol{\zeta}^{(k)},\boldsymbol{l}^{(k)})},\nonumber \\
    &\zeta_n^{(k)}=\frac{d_{n,1}^{(k)}-{d_{n,1}^{'(k)}}}{2[\frac{w_1}{f_n^{s(k)}}\!\!+\!\!w_2k^S(f_n^{s(k)})^2]}. \nonumber
\end{align}

For the last function $V_n(\boldsymbol{x},\boldsymbol{d})$, it can be similarly transformed to $\widetilde{V}_n(\boldsymbol{x},\boldsymbol{d}|\boldsymbol{\eta}^{(k)},\boldsymbol{\theta}^{(k)},\boldsymbol{q}^{(k)},\boldsymbol{z}^{(k)})$ where $\boldsymbol{\eta}$ and $\boldsymbol{\theta}$ are auxiliary variables, and $\boldsymbol{q}$ and $\boldsymbol{z}$ are the constants. Then the details of $\widetilde{V}_n$ is: $\widetilde{V}_n=[d_{n,1}^{'2}(\eta_n^{(k)}+q_n^{(k)})+\frac{x_{n,1}^2}{4(\eta_n^{(k)}+q_n^{(k)})}]C+[d_{n,2}^2(\theta_n^{(k)}+z_n^{(k)})+\frac{x_{n,2}^2}{4(\theta_n^{(k)}+z_n^{(k)})}](C-\frac{w_1+w_2\Bar{P}}{r_0})$. When auxiliary variables are not equal to zero, auxiliary variables are expressed as:
\begin{align}
    \eta_n^{(k)}=\frac{x_n^{(k)}}{2d_{n,1}^{'(k)}}, ~~\theta_n^{(k)}=\frac{x_{n,2}^{(k)}}{2d_{n,2}^{(k)}}.\nonumber
\end{align}

Until now, the optimization variables in the term $w_1T_n+w_2E_n$ of the objective function in (\ref{ofSCA}) are transformed into a convex function by using the updated transform. $\hfill\blacksquare$

}

% \newpage

 

% \end{IEEEbiography}



% \vfill

\end{document}